\documentclass[10pt,twocolumn,final]{IEEEtran}
\usepackage[dvips,final]{graphicx}
\usepackage{latexsym}
\usepackage{amssymb}
\usepackage[cmex10]{amsmath}
\usepackage{amsthm}
\usepackage{bm}
\usepackage{bbm}
\usepackage{array}
\usepackage{balance}
\usepackage[caption=false,font=footnotesize]{subfig}
\usepackage{cite}
\usepackage{algorithmic}
\usepackage{algorithm}
\usepackage{multirow}
\usepackage{pgfplots} 

\pgfplotsset{compat=newest}
\pgfplotsset{plot coordinates/math parser=false}

\newcommand{\argmax}{\operatornamewithlimits{arg\,max}}

\title{Content-Aware Delivery of Scalable Video in Network Coding Enabled Named Data Networks}

\author{Eirina~Bourtsoulatze,
	    Nikolaos~Thomos, 
	    Jonnahtan Saltarin, 
        and~Torsten~Braun %
\thanks{E. Bourtsoulatze, J. Saltarin and T. Braun are with the Communication and Distributed Systems laboratory (CDS), Institute for Computer Science and Applied Mathematics, Bern, Switzerland (e-mail: bourtsoulatze@iam.unibe.ch; saltarin@iam.unibe.ch; braun@iam.unibe.ch). N. Thomos is with the University of Essex (e-mail: thomos@essex.ac.uk).}%
}

\begin{document}
\maketitle

\begin{abstract}
In this paper, we propose a novel network coding enabled NDN architecture for the delivery of scalable video. Our scheme utilizes network coding in order to address the problem that arises in the original NDN protocol, where optimal use of the bandwidth and caching resources necessitates the coordination of the forwarding decisions. To optimize the performance of the proposed network coding based NDN protocol and render it appropriate for transmission of scalable video, we devise a novel rate allocation algorithm that decides on the optimal rates of Interest messages sent by clients and intermediate nodes. This algorithm guarantees that the achieved flow of Data objects will maximize the average quality of the video delivered to the client population. To support the handling of Interest messages and Data objects when intermediate nodes perform network coding, we modify the standard NDN protocol and introduce the use of Bloom filters, which store efficiently additional information about the Interest messages and Data objects. The proposed architecture is evaluated for transmission of scalable video over PlanetLab topologies. The evaluation shows that the proposed scheme performs very close to the optimal performance. 
\end{abstract}


\section{Introduction}
\label{sec:intro}

During the last decade, we have witnessed a radical change of the video production and communication model. Besides their traditional role as content consumers, users nowadays are often able to produce and share their own video content. This change has been fostered by the emergence of affordable price camera enabled mobile devices with various capabilities, and has resulted in video data dominating the overall
Internet traffic \cite{Cisco:15}. To address users' heterogeneity in terms of display capabilities, processing power, network connectivity, \emph{etc}., the video is often encoded in multiple qualities and resolutions \cite{SVC2005}. This enables users to access the video of their interest encoded in a quality and resolution that matches the capabilities of their device. At the same time, however, this necessitates efficient mechanisms that can enable the delivery of scalable data to heterogeneous clients.

The Internet protocol, designed originally for delay-tolerant applications, such as messaging and file downloading, fails to deal efficiently with the growing volume of time sensitive video traffic.  The introduction of protocols like TCP, DASH \cite{SodagarMM2011,StockhammerMMSys2011} and RTP/RTSP have permitted to partly handle the real-time and on demand delivery of large volumes of video traffic. However, these solutions have rendered the network operation and management complex. To cope with the inefficiencies of the IP host-centric communication model such as scalability, mobility, \textit{etc.}, Information-Centric Network (ICN) architectures  \cite{XylomenosComSurvey2014} have been proposed as an alternative solution. The ICN communication paradigm focuses on the name of the content rather than on its location. Thus, the content is searched by its name and can be retrieved from any location where it may be permanently or temporarily stored without the need to establish multiple dedicated server-client connections. This content-centric approach makes use of the available network caching capacity and reduces the redundancy of the transmitted content. 

Among the existing ICN architectures, Named Data Networking (NDN) \cite{JacobsonCoNEXT2009} has gained significant popularity because of its intuitive naming scheme and the way content requests and data forwarding are handled. The NDN model is receiver driven. The receiver (client) initiates the content delivery by sending Interest messages with the name identifier of the requested content to some (or all) of its outgoing faces. Once an Interest message reaches an uplink node, the node's cache is searched for a matching Data object. If the requested data exists, the node sends the data backwards on the requesting face. If there is an indication that the data will be available at a later time instant, the Interest is kept at the node and consumed later when the data arrives. Otherwise, the Interest is forwarded to other nodes according to the employed forwarding strategy. The Data objects that travel towards the end users can be cached in the intermediate nodes and can be used later to consume new requests for the same content. 

Although solutions exist that permit to deal with real-time and on-demand video delivery, it is generally acknowledged that NDN and in overall ICN architectures are not yet video ready \cite{TsilopoulosPV2013}. For example, in live streaming the main bottleneck of NDN is that each content packet is independently requested by Interest messages, which increases the network load and raises scalability issues. To address this problem, Interest aggregation \cite{ByunICC2013} and persistent Interest packets \cite{TsilopoulosACMICN2011} can be used. However, the use of such approaches is not trivial, as the loss of a single Interest message can lead to the  loss of multiple Data objects. In Video-on-Demand systems, the main problem of using ICN arises from the fact that there are no reliable estimations of the available end-to-end bandwidth. Adaptive video streaming over ICN is achieved by deploying the DASH protocol over NDN \cite{LiuICC2013}. This is driven by the conceptual similarities of the NDN and DASH protocols. The use of the DASH protocol results in significant performance gains and allows the use of the multiple interfaces of the devices.

The requirement that each packet of a data stream should be requested explicitly by sending an Interest message is relaxed by equipping the NDN protocol with network coding capabilities \cite{MontpetitNoM2012}. Network coding \cite{AhlswedeTIT2000} can improve the use of the network resources \cite{ThomosTMM2011}, enhance the resilience of the communication, simplify the scheduling, remove the need for coordination, \textit{etc.} With network coding the intermediate network nodes linearly combine the received packets prior to forwarding them to the outgoing links. The linear combinations are performed in a Galois field. To deploy network coding in practical settings, Randomized Linear Network Coding (RLNC) \cite{HoAllerton2003} has been proposed. The network coded packets have a header that contains  the network coding coefficients and permits the end users to decode the linearly combined packets. The introduction of the concept of generations \cite{ChouAllerton2003} helps to limit the overhead information carried by each network coded packet and renders it appropriate for video transmission.

The potential of network coding enabled NDN protocols in \cite{MontpetitNoM2012} motivated researchers to explore the use of coding enabled NDN variants. CodingCache \cite{WuACMICN2013} has focused on the caching problems and has shown that cache diversity due to network coding helps to increase the cache hit rate. The solution in \cite{MontpetitNoM2012} has been analytically studied in \cite{LloreaICC2013} for the butterfly network. Other coding solutions such as Raptor codes have been examined in \cite{AnastasiadesICC2015} where RC-NDN, a variant of the NDN, is presented. This scheme shows the benefits of using Raptor codes in mobile networks. 

In this paper, we propose a content aware video delivery scheme for network coding enabled NDN architectures. We focus on the transmission of scalable video content \cite{SVC2005} in order to deal with the requests of users that have diverse demands in terms of the video quality they want to receive. We employ  prioritized random linear network coding (PRLNC) \cite{ThomosTMM2011, KurdogluICME2011} in order to respect the unequal importance of the video layers. PRLNC is applied in an embedded way forming packets that belong to classes of decreasing significance. We first derive the optimal rate allocation for the transmission of Interest messages in order to achieve the flow of Data objects that maximizes the average video quality in the client population. We then present in detail the new features of our network coding enabled NDN architecture. These new features include the appropriate naming scheme and the processing functions that permit to handle both the Interest messages and the Data objects when the intermediate nodes perform network coding. In order to deal with the ambiguity that arises from the use of content names that do not specify unique Data objects but rather a set of network coded packets that belong to the same class and generation, we propose the use of Bloom filters that compactly store additional information about the Interest messages and Data objects. Finally, we design the optimal content-aware forwarding strategy based on the solution of the rate allocation problem. The forwarding strategy guarantees that a sufficient number of Interest messages will be optimally forwarded so that the innovative rate of Data objects remains sufficiently high. We evaluate the performance of the presented scheme for scalable video transmission with respect to the experienced video quality. The evaluation shows that the proposed method results in close to optimal performance in terms of the achieved video quality.

In summary, the main contributions in this paper are the following:
\begin{itemize}

\item
we propose a novel content-aware network coding enabled NDN architecture appropriate for delivering layered data in general and scalable video in particular;

\item
we formulate the optimal flow rate allocation problem for the transmission of Interest messages so that the achieved rate of Data objects maximizes the average video quality over the client population;

\item 
we redesign the functions that handle the Interest messages and Data objects in order to enable the processing of network coded content;

\item
we propose the use of Bloom filters in order to resolve problems related to the introduction of network coding in the NDN architecture;

\item 
we design an optimized forwarding strategy that handles the forwarding of Interest messages based on the optimal solution of the flow rate allocation problem.

\end{itemize}

The rest of the paper is organized as follows. In Section \ref{sec:system}, we present the overview of the system and discuss the problem of optimally delivering layered video in our setting. Next, in Section  \ref{sec:rateallocation}, we formulate the flow rate allocation problem for the transmission of Interest messages and present the subgradient based optimization algorithm for obtaining the optimal solution. Section \ref{sec:implementation} discusses the design of the network coding enabled NDN protocol. The performance evaluation of the proposed protocol is presented in Section \ref{sec:results}. Finally, Section \ref{sec:discussion} summarizes the paper.

\section{System description}
\label{sec:system}

\subsection{Network model}
\label{sec:networkmodel}

We consider a multi-hop wireline network with full duplex links that can simultaneously carry information in both directions between pairs of connected nodes. The network is modeled by a directed graph $\mathcal{G} = (\mathcal{V},\mathcal{E})$, where $\mathcal{V}$ and $\mathcal{E}$ denote the set of network nodes and network links, respectively. For notational convenience, we assume that every physical link in the network is modelled by a pair of directed links in the graph $\mathcal{G}$, such that $(i,j)\in \mathcal{E}$ if and only if $(j,i) \in \mathcal{E}$, where the tuple $(i,j)$ denotes the directed link from node $i \in \mathcal{V}$ to node $j \in \mathcal{V}$. Assuming this convention, the set of network links $\mathcal{E}$ can be written as the union of two disjoint sets $\mathcal{E}_I$ and $\mathcal{E}_D$, \emph{i.e.}, $\mathcal{E} = \mathcal{E}_I\cup \mathcal{E}_D$, such that $(i,j) \in \mathcal{E}_I$ if and only if $(j,i)\in \mathcal{E}_D$. We further assume that the directed graphs $\mathcal{G}_I = (\mathcal{V}, \mathcal{E}_I)$ and $\mathcal{G}_D = (\mathcal{V}, \mathcal{E}_D)$ are acyclic which ensures that the information does not cycle within the network. Each pair of links that corresponds to a single physical duplex channel is characterized by the cumulative transmission rate $B_{ij}$ in bits per second (bps) that can be allocated proportionally to the traffic load in each direction.

The set of network nodes $\mathcal{V}$ consists of a server node $s$, a set of intermediate nodes $\mathcal{I}$ and a set of client nodes $\mathcal{U}$.   We have $\mathcal{V} = s \cup \mathcal{I} \cup \mathcal{U}$. The server generates video content which is subsequently delivered to the clients through the intermediate nodes following the NDN protocol. The video delivery is initiated by the clients that transmit Interest messages for the desired video content. The Interest messages are routed towards the server according to the routing information stored in the intermediate nodes' Forwarding Information Base tables until a matching Data object is found. The Data object is then transmitted back to the client following the reverse path of that followed by the Interest message.

\subsection{Prioritized delivery of scalable video}
\label{sec:prioritizedNC}

Due to the heterogeneity of the network clients in terms of bandwidth resources and video display capabilities, the clients may request video content encoded at different bit rates so as to better adapt the quality of the delivered video to the available resources. In order to meet the diverse clients' demands, the video server $s$ encodes the video progressively into $L$ video layers with the scalable extension (SVC) of the H.264/AVC compression standard \cite{SVC2005}. The $l$-th video layer is encoded at rate $R_l$ expressed in packets per second. The video layers include the base layer, which provides the basic video quality, and $L-1$ enhancement layers, which offer an incremental improvement of the video quality. According to the SVC standard, the $l$-th video layer can be decoded only if all the previous video layers have been successfully decoded. The decoding dependencies between the video layers define a hierarchical structure with the base layer being assigned the highest level of importance, while each subsequent layer has a decreasing degree of importance with respect to the previous layers.

In order to improve the network performance in terms of throughput and efficient use of available resources, the server and the intermediate nodes  combine the data packets with Prioritized Random Linear Network Coding (PRLNC) \cite{ThomosTMM2011}. To enable the network coding operations without incurring an additional decoding delay penalty \cite{ChouAllerton2003}, the source data is segmented into generations. Each generation comprises source video packets with similar decoding deadlines. We consider that the data of the $l$-th layer in each generation is packetized into $\alpha_l$ packets. Prior to transmission, the packets within each generation are encoded by means of PRLNC \cite{ThomosTMM2011}. Specifically, the packets are first categorized into $L$ classes of decreasing importance where the $l$-th class consists of source packets that belong to the first $l$ layers. The number of source packets that belong to the $l$-th class is $\beta_l = \sum_{k = 0}^l{\alpha_k}$, while class $L-1$ contains $\beta_{L-1} = \sum_{k = 0}^{L-1}{\alpha_k}$ packets and thus includes all the packets of a generation. Packets within each class are then encoded with random linear coding. The coded packet that results from randomly combining the source packets from class $l$ is hereafter referred to as network coded packet of class $l$. 

The decoding of the network coded packets is done only at the client nodes with the help of Gaussian elimination upon receiving a sufficient number of network coded packets. A client can decode a generation of the $l$-th video layer only upon receiving $\beta _l$ innovative packets of this generation from classes $0, 1,\dots, l$. Note that class $l$ contains up to $\beta_l$ innovative network coded packets. The PRLNC method respects the intrinsic prioritized structure of the video data by assigning higher importance to layers with smaller index. It thus avoids penalizing clients that do not have sufficient resources to jointly decode all the video layers and offers an adaptive data delivery solution for clients with heterogeneous resources.

\subsection{Optimal delivery of SVC video in NDN}
\label{sec:generalproblemformulation}

In our network coding enabled NDN architecture for scalable video delivery, we adopt the convention that every Data object represents a single network coded packet of some class $l$ and generation $g$. The Interest messages express a request for a network coded packet of a certain class and generation, and can be consumed by any network coded packet of the specified class and generation. We assume that the clients generate Interest messages at a constant rate. In this context, our goal is to ensure a constant streaming bit rate and playback quality at each client.

The time constrained nature of the video imposes strict requirements on the video delivery deadlines. An optimized forwarding strategy is therefore essential in order to achieve the delivery deadlines set by the video application. In this work, we propose to design the forwarding strategy based on the optimal rate allocation for Interest messages that maximizes the average quality of the video delivered to the client population.

Let $\bm{f} = \{f^0, f^1, \dots, f^{L-1}\}$ be a set of flows, where $f^l$, $\forall l \in \mathcal{L}$, is the flow of Data objects of class $l$ in the network and $\mathcal{L} = \{0,1,\dots,L-1\}$ is the set of packet classes. The average video quality in the client population can be written as 
\begin{equation} 
	\overline{Q}(\bm{f}) = \frac{1}{U}\sum_{u \in \mathcal{U} }Q_{u}(\bm{f})
\label{eq:objective1}
\end{equation}
where $U = |\mathcal{U}|$ is the number of clients in the network and  $Q_u(\bm{f})$ is the quality of the video delivered to client $u$ as a function of the flows $f^l$ of the Data objects. The function $Q_u(\bm{f}) $ is a piecewise constant function. It can be expressed as a linear combination of indicator functions as 
\begin{equation}
	Q_u(\bm{f}) = \sum_{l = 0}^ {L-1} (q_l-q_{l-1}) \mathbbm{1}_l(\bm{f})
\label{eq:qualityfunction}
\end{equation}
where $q_l$ is the video quality achieved after decoding the $l$-th video layer with $q_{-1} = 0$. The indicator function $\mathbbm{1}_l(\bm{f})$ is defined as 
\begin{equation}
	\mathbbm{1}_l(\bm{f}) = \begin{cases}
	1, &   \parbox[t]{.5\textwidth}{if the $l$-th video layer can be decoded given $\bm{f}$}\\
	0, &   \mbox{otherwise}
	\end{cases}
\label{eq:indicator}
\end{equation}
Given that our objective is to maximize the average video quality in \eqref{eq:objective1}, the problem boils down to (i) determining the rate of the Interest messages for each class of packets that must be transmitted from the clients, and (ii) designing the optimal forwarding strategy at clients and intermediate nodes, in order to achieve the set of flows $\{f^0, f^1, \dots, f^{L-1}\}$ that maximize the average video quality.

\section{Optimal content-aware flow allocation}
\label{sec:rateallocation}

We now present our algorithm for determining the optimal forwarding strategy at the clients and at the intermediate nodes in a network coding enabled NDN architecture for the delivery of scalable video. We derive the forwarding strategy by casting this problem in a rate allocation problem that aims at maximizing the average video quality in the client population. Our approach relies on the observation that Interest messages generated by different clients and expressing interest for a network coded packet of the same class $l$ and generation $g$ can be aggregated upon arriving at a node and only a single Interest message for a network coded packet of this class and generation has to be transmitted further towards the server to fetch the Data object that will consume all the aggregated Interest messages. Taking into account this property, we can make use of the concept of \emph{conceptual flows} presented in \cite{LiINFOCOM05} to design our content-aware algorithm for optimal forwarding of Interest messages. In particular, the overall flow of Interest messages from the clients to the servers can be regarded as consisting of $U$ unicast conceptual flows from each client to the server, where $U =|\mathcal{U}|$ is the number of clients in the network. Conceptual flows are network flows that can co-exist in the network without contending for the link bandwidth. Therefore, the rate of the actual flow on a link is the maximum of the rates of all the conceptual flows that pass through this link.     

\subsection{Rate allocation problem}
\label{sec:rateallocproblem}

Let $r^{u,l}$ denote the conceptual flow of Interest messages expressing a request for network coded packets of class $l$ and originating from client $u \in \mathcal{U}$, and let $r_{ij}^{u,l}$ be the rate of the conceptual flow $r^{u,l}$ on link $(i,j)$, $\forall (i,j) \in \mathcal{E}_I$. The rate of the actual flow of Interest packets for class $l$ network coded packets on link $(i,j)$ will be $z_{ij}^l = \max_{u\in\mathcal{U}} r_{ij}^{u,l}$. Let $f_{ji}^l$ denote the rate of the flow of Data objects of class $l$ on the link $(j,i) \in \mathcal{E}_D$. Since every Interest message transmitted on link $(i,j) \in \mathcal{E}_I$ is eventually consumed by a matching Data object transmitted on link $(j,i) \in \mathcal{E}_D$, the rate of the flow of Data objects  on link $(j,i)$ is equal to the rate of the flow of Interest messages $z_{ij}^{l}$ on link $(i,j)$, \emph{i.e.}, we have $z_{ij}^{l} = f_{ji} ^{l}$, $\forall (i,j) \in \mathcal{E}_I$ and $(j,i) \in \mathcal{E}_D$.

Recall from Section \ref{sec:generalproblemformulation} that the average video quality function $\overline{Q}(\bm{f})$ is a sum of piecewise constant functions, and is therefore also a piecewise constant function. That means that there can be multiple sets of flow values that maximize the function in \eqref{eq:objective1}. In order to resolve this ambiguity and give higher priority to more important classes of packets, we introduce a flow cost function for each user $u$ that represents the overall cost of requesting packets of different classes:
\begin{equation}
	\begin{aligned}
		C_u(\bm{r}^u)  = \bm{c}^T \bm{r}^u  =  \sum_{l = 0}^{L-1} c_l \sum_{j:(u,j)\in\mathcal{E}_I}r_{uj}^{u,l}
	\end{aligned}
	\label{eq:costfunction}
\end{equation}
where $\bm{c} = (c_0, c_1, \dots, c_{L-1})^T$ is the cost vector, $\bm{r}^u = (r^{u,0}, r^{u,1}, \dots, r^{u,L-1})^T$ is the vector of conceptual flows of different classes and $r^{u,l} = \sum_{j:(u,j)\in\mathcal{E}_I}r_{uj}^{u,l}$, $\forall u \in \mathcal{U}$, due to the flow conservation property. $c_l$ is the cost for requesting network coded packets of class $l$. The costs can be chosen arbitrarily, and must satisfy the following constraints:
\begin{subequations}
\begin{align}
&0 < c_0 < c_1 <\dots < c_{L-1} \label{eq:costconstraint1}\\
&c_l < \frac{q_l - q_{l-1}}{\sum_{k=0}^l R_k}, \; l = 1, 2,,\dots, L-1\label{eq:costconstraint2}
\end{align}
\end{subequations}
Constraint \eqref{eq:costconstraint1} implies that lower cost is assigned to more important packet classes, while higher cost is assigned to less important packet classes. This guarantees that more important classes are given higher priority in the rate allocation algorithm compared to the less important classes. Constraint \eqref{eq:costconstraint2} ensures that the introduction of the cost function into the optimization problem does not alter the value of the objective function in \eqref{eq:objective1} at the optimal rate allocation solution. 

By including the cost function in \eqref{eq:costfunction} in the problem formulation, we can cast our rate allocation problem as an optimization problem that seeks to maximize the average video quality while minimizing the average flow cost. Formally, this optimization problem can be written as:
\begin{equation}
	\argmax_{\bm{f}, \bm{r}^u} \frac{1}{U} \sum_{u \in \mathcal{U}}\Big{(} Q_u(\bm{f}) - C_u(\bm{r}^u)\Big{)}
	\label{eq:optimization1}
\end{equation}
Taking into account that the quality of the delivered video at client $u$ depends only on the rate of the flow of Data objects on the input links of the client, the video quality at user $u$ can be expressed as a function of the rates of the conceptual flows of Interest messages as follows:  
\begin{equation}
\small
\begin{aligned}
	Q_u(\bm{f}) &= Q_u(\sum_{j:(j,u) \in {\mathcal{E}_D}}f_{ju}^0,\sum_{j:(j,u) \in {\mathcal{E}_D}}f_{ju}^1,\dots,\sum_{j:(j,u) \in {\mathcal{E}_D}}f_{ju}^{L-1} ) \\
	& = Q_u(\sum_{j:(u,j) \in {\mathcal{E}_I}}r_{uj}^{u,0},\sum_{j:(u,j) \in {\mathcal{E}_I}}r_{uj}^{u,1},\dots,\sum_{j:(u,j) \in {\mathcal{E}_I}}r_{uj}^{u,L-1} ) \\
	& = Q_u(\bm{r}^u)
\end{aligned}
\label{eq:objective2}
\end{equation}
where we have used the fact that  $f_{ju}^l = z_{uj}^l $ and $z_{uj}^l = \max_{w \in \mathcal{U}}r_{wj}^{w,l} = r_{uj}^{u,l}$. By combining \eqref{eq:optimization1}, \eqref{eq:costfunction} and \eqref{eq:objective2}, the objective function in the optimization problem in \eqref{eq:optimization1} can be rewritten as:
\begin{equation}
	\argmax_{\bm{r} \in \mathcal{R}} \frac{1}{U} \sum_{u \in \mathcal{U}} \Big{(}Q_u(\bm{r}^u) - \bm{c}^T\bm{r}^u\Big{)
} 
	\label{eq:optimization2}
\end{equation}
where $\bm{r}$ is the vector of all rate variables $r_{ij}^{u,l}$, $\forall (i,j) \in \mathcal{E}_I, \; \forall u \in \mathcal{U}, \; \forall  l \in \mathcal{L}$.  The polytope $\mathcal{R}$ is defined by the following set of linear equality and inequality constraints:
\begin{subequations}
\small
	\begin{align}
		&r_{ij}^{u,l} \geq 0, \quad  \forall u \in \mathcal{U}, \;\forall (i,j) \in \mathcal{E}_I, \; \forall l \in \mathcal{L} \label{eq:constraint1}\\
		&\sum_{n: (n,i) \in \mathcal{E}_I} r_{ni}^{u,l} = \sum_{j:(i,j)\in \mathcal{E}_I} r_{ij}^{u,l}, \quad \forall u \in \mathcal{U},\;\forall i\in\mathcal{I}, \;  \forall l \in \mathcal{L}\label{eq:constraint2}\\
		&\sum_{k = 0}^l \sum_{j:(u,j)\in \mathcal{E}_I} r_{uj}^{u,k} \leq \sum_{k = 0}^l R_k, \quad \forall u \in \mathcal{U}, \; \forall l\in \mathcal{L} \label{eq:constraint3}\\
		&x_{ij}^l \geq r_{ij}^{u,l}, \quad \forall u \in \mathcal{U}, \; \forall (i,j) \in \mathcal{E}_I, \; \forall l \in \mathcal{L} \label{eq:constraint4}\\
		&\sum_{l \in \mathcal{L}} x_{ij}^l(p_I + p_D) \leq B_{ij} \quad\forall (i,j) \in \mathcal{E}_I\label{eq:constraint5}
	\end{align} 
	\label{eq:optconstraints}
\end{subequations}
where $p_I$ and $p_D$ in \eqref{eq:constraint5} are the size (in bits) of the Interest messages and Data objects, respectively. The objective function in \eqref{eq:optimization2} is a function of only the rates of the conceptual flows of Interest messages on the input links of the clients. The set of constraints \eqref{eq:constraint1} -\eqref{eq:constraint5} defines the set of feasible rate allocations for the conceptual flows of Interest messages. Specifically, constraint \eqref{eq:constraint2} is the flow conservation constraint that holds for every conceptual flow of Interest messages, since every conceptual flow is a unicast flow from the client to the server. This constraint guarantees that the number of Interest messages that enter an intermediate node is equal to the number of Interest messages that are transmitted from the node for every conceptual flow and for every class of network coded packets. Constraint \eqref{eq:constraint3} is the innovative\footnote{A network coded packet is considered innovative with respect to a set of network coded packets, when it cannot be generated by linearly combining the packets in the set.} rate constraint which states that the rate of Interest messages generated by a client for each class of network coded packets should not exceed the maximum rate of innovative Data objects that can be provided by the source. Constraints \eqref{eq:constraint2} and \eqref{eq:constraint3} together ensure that the clients do not request Data objects at a rate which is higher than the rate at which innovative Data objects are generated by the server. The variable $x_{ij}^l$ in \eqref{eq:constraint4} is an auxiliary variable that represents the actual transmission rate of Interest messages for network coded packets of class $l$ on link $(i,j)$. It upper bounds the transmission rates of the conceptual flows of Interest messages on every link. The actual rates $x_{ij}^l$ are further bounded by the available bandwidth on the link which is captured by the constraints in \eqref{eq:constraint5}. Constraint \eqref{eq:constraint5} states that the bandwidth required to transmit the Interest messages and the Data objects that consume these Interest messages on a link, should not exceed the overall bandwidth available for that link.

\subsection{Subgradient-based algorithm}
\label{sec:algorithm}

In this section, we present an efficient algorithm based on Lagrangian relaxation and the subgradient method for solving the optimization problem formulated in \eqref{eq:optimization2}. As previously discussed, the video quality function in \eqref{eq:objective2} is a piecewise constant function and therefore, the objective function in \eqref{eq:optimization2} is piecewise affine and thus also piecewise concave. 

We first relax the coupling constraints in \eqref{eq:constraint4} and obtain the Lagrangian dual function. This choice is driven by fact that the resulting Lagrangian dual problem can be decomposed into several optimization subproblems that can be solved independently. The Lagrangian dual function is given by
\begin{equation}
\small
 	\begin{aligned}
		L(\bm{\mu})& = \max_{\mathcal{R}^\prime} \frac{1}{U} \sum_{u \in \mathcal{U}} \Big{(}Q_u(\bm{r}^u)- \bm{c}^T\bm{r}^u\Big{)}\\
		& - \sum_{u\in\mathcal{U}} \sum_{(i,j)\in \mathcal{E}_I} \sum_{l \in \mathcal{L}} \mu_{ij}^{u,l}(r_{ij}^{u,l} - x_{ij}^l)\\
		&= \max_{\mathcal{R}^\prime }\sum_{u \in \mathcal{U}} \Big{(}\frac{1}{U} (Q_{u}(\bm{r}^u) - \bm{c}^T\bm{r}^u)
		 - \sum_{(i,j)\in\mathcal{E}_I}\sum_{l\in\mathcal{L}}\mu_{ij}^{u,l}r_{i,j}^{u,l} \Big{)}\\
	 	& + \sum_{(i,j)\in\mathcal{E}_I}\sum_{l\in\mathcal{L}}\Big{(} \sum_{u\in\mathcal{U}}\mu_{ij}^{u,l}\Big{)}x_{ij}^l
	\label{eq:lagrangian}
	\end{aligned} 
\end{equation}
where $\mathcal{R}^\prime$ is the polytope defined by the linear constraints \eqref{eq:constraint1}, \eqref{eq:constraint2}, \eqref{eq:constraint3} and \eqref{eq:constraint5}. The Lagrangian dual of the primal problem in \eqref{eq:optimization2} is then 
\begin{equation}
	\min_{\bm{\mu} \geq 0} L(\bm{\mu})
	\label{eq:dualproblem}
\end{equation}
where $\bm{\mu}$ is the vector of Lagrangian multipliers $\mu_{ij}^{u,l}$. We can observe that the Lagrangian subproblem in \eqref{eq:lagrangian} can be decomposed into several optimization subproblems that can be solved independently. These optimization subproblems consist of $U$ maximization problems 
\begin{equation}
\small
\begin{aligned}
	\argmax _{\bm{r} \in \mathcal{R}_u} \frac{1}{U} \Big{(}Q_u(\bm{r}^u) - \bm{c}^T\bm{r}^u \Big{)} -  \sum_{(i,j)\in\mathcal{E}_I}\sum_{l\in\mathcal{L}}\mu_{ij}^{u,l}r_{i,j}^{u,l}, \quad
	\forall u \in \mathcal{U}
	\end{aligned}
	\label{eq:subproblem1}
\end{equation}
where the polytope $\mathcal{R}_u$ is defined for every user $u$ as 
\begin{subequations}
\small
	\begin{align}
		r_{ij}^{u,l} \geq 0, &\quad \forall (i,j) \in \mathcal{E}_I, \; \forall l \in \mathcal{L} \label{eq:dualconstraint1}\\
		\sum_{n: (n,i) \in \mathcal{E}_I} r_{ni}^{u,l} = \sum_{j:(i,j)\in \mathcal{E}_I} r_{ij}^{u,l}, &\quad\forall i\in\mathcal{I}, \;  \forall l \in \mathcal{L}\label{eq:dualconstraint2}\\
		\sum_{k = 0}^l \sum_{j:(u,j)\in \mathcal{E}_I} r_{uj}^{u,k} \leq \sum_{k = 0}^l R_k, &\quad  \forall l\in \mathcal{L} \label{eq:dualconstraint3}
	\end{align}
\end{subequations}
and $|\mathcal{E}_I|$ maximization problems
\begin{equation}
\small
\begin{aligned}
\argmax_{\bm{x}} &\sum_{l\in\mathcal{L}}\Big{(} \sum_{u\in\mathcal{U}}\mu_{ij}^{u,l}\Big{)}x_{ij}^l, \quad \forall (i,j) \in \mathcal{E}_I\\
\mbox{s.t.} &\quad \sum_{l \in \mathcal{L}} x_{ij}^l(p_I + p_D) \leq B_{ij}
\end{aligned}
\label{eq:subproblem2}
\end{equation}
where $\bm{x}$ is the vector of the rates $x_{ij}^l$, $\forall (i,j) \in \mathcal{E}_I$ and $\forall l \in \mathcal{L}$, of the actual flows of Interest messages. The maximization problems in \eqref{eq:subproblem2} are linear programs that can be solved with one of the general LP optimization methods, such as the Simplex algorithm. The optimization problem in \eqref{eq:subproblem1} consists in maximizing a piecewise linear objective function. Since it is hard to directly optimize the objective function in \eqref{eq:subproblem1}, the optimization problem in \eqref{eq:subproblem1} can be further transformed into a two level optimization problem:
\begin{equation}
	\max_{k \in \mathcal{L}}\max_{\mathcal{R}_u \cap \mathcal{R}_k} \frac{1}{U} \Big{(}q_k - \bm{c}^T\bm{r}^u \Big{)}  -  \sum_{(i,j)\in\mathcal{E}_I}\sum_{l\in\mathcal{L}}\mu_{ij}^{u,l}r_{i,j}^{u,l}, \quad \forall u \in \mathcal{U}
	\label{eq:subproblem2modified}
\end{equation}
where $\mathcal{R}_u$ is defined in \eqref{eq:dualconstraint1}-\eqref{eq:dualconstraint3}. The polytope $\mathcal{R}_k$ defines the rate region where the class $k$ packets are decodable and is given by the following set of linear constraints 
\begin{subequations}
	\begin{align}
		&\sum_{m = 1}^l \sum_{j : (u,j) \in \mathcal{E}_I} r_{uj}^{u,m} \leq \sum_{m =1}^l R_m, \; \forall l \in \{0, \dots, k-1\} \label{eq:dualconstraint5}\\
		&\sum_{m = 1}^k \sum_{j : (u,j) \in \mathcal{E}_I} r_{uj}^{u,m} = \sum_{m =1}^k R_m \label{eq:dualconstraint6}\\
		& \sum_{m = 1}^l \sum_{j : (u,j) \in \mathcal{E}_I} r_{uj}^{u,m} < \sum_{m =1}^l R_m, \; \forall l \in \{k+1,\dots, L\}\label{eq:dualconstraint7}
	\end{align}
\end{subequations}
At the lower level of the optimization problem in \eqref{eq:subproblem2modified}, the objective function is linear and is maximized over the rate region where the $k$-th class of network coded packets is decodable. Constraints \eqref{eq:dualconstraint5} and \eqref{eq:dualconstraint6} guarantee that the rate of the Data objects that will be delivered to the client is sufficient to decode network coded packets of class $k$, while at the same time the rate of Data objects per each class does not exceed the available innovative rate for this class of packets. Constraint \eqref{eq:dualconstraint7} ensures that the rate of Data objects delivered to the client is not sufficient to decode network coded packets of classes higher than $k$. At the higher level of the optimization problem in \eqref{eq:subproblem2modified}, a pointwise maximization with respect to the packet classes is performed by selecting the solution that yields the best value of the objective function. 

To solve the Lagrangian dual problem in \eqref{eq:dualproblem}, we apply the subgradient algorithm. Specifically, we select an initial set of non-negative values of Lagrangian multipliers $\mu_{ij}^{u,l}[0]$, $\forall (i,j) \in \mathcal{E}_I, \; \forall u \in \mathcal{U}, \; \forall l \in \mathcal{L}$, and update them at every iteration $t =  1, 2, \dots$ according to the following rule:
\begin{eqnarray}
	\mu _{ij} ^{u,l}[t] =  & \max \{0, \mu_{ij}^{u,l}[t-1] +\theta[t](r_{ij}^{u,l}[t] - x_{ij}^{u,l}[t])\}, 
	\\ & \quad \forall (i,j) \in \mathcal{E}_I, \; \forall u \in \mathcal{U}, \; \forall l \in \mathcal{L} \nonumber
\label{eq:lagrangeupdate}
\end{eqnarray}
where $r_{ij}^{u,l}[t]$ and $x_{ij}^{u,l}[t]$ are the solutions to the optimization problems in \eqref{eq:subproblem1} and \eqref{eq:subproblem2} at the $t$-th iteration of the subgradient algorithm given the current value of the Lagrangian multipliers $\mu_{ij}^{u,l}[t]$. The step size $\theta[t]$ controls the convergence properties of the subgradient algorithm. When the step size is chosen such that 
\begin{equation}
	\theta[t] > 0, \quad \lim_{t\rightarrow \infty} \theta[t] = 0, \quad \sum_{t=1}^\infty\theta[t] = \infty 
\label{eq:stepsize}
\end{equation}
the subgradient algorithm is guaranteed to converge to the optimal solution of the Lagrangian dual problem in \eqref{eq:dualproblem}\cite{Sherali96}. Here we choose $\theta[t] = {a}/{(b+ct)}$, $\forall t$, with $a >0$, $b \geq 0$ and $c >0$, which satisfies the conditions in \eqref{eq:stepsize}. In order to recover the optimal solution to the primal problem in \eqref{eq:optimization2}, we use the method introduced by Sherali \emph{et al.} \cite{Sherali96}. Specifically, at every iteration $t$, the value of the primal variables $r_{ij}^{u,l}$ and $x_{ij}^{l}$ is constructed according to the following update rule:
\begin{subequations}
	\begin{align}
		\hat{r}_{ij}^{u,l}[t] = \sum_{h = 1}^t \nu_h [t]r_{ij}^{u,l}[h] \label{eq:primalrecovery1}\\
		\hat{x}_{ij}^{l}[t] = \sum_{h = 1}^t \nu_h [t]x_{ij}^{l}[h]\label{eq:primalrecovery2}
	\end{align}
\end{subequations}
where $\sum_{h = 1}^t \nu_h[t] = 1$  aned $\nu_h[t] \geq 0$, for $ h = 0,1,\dots, t$. A valid choice for the convex combination weights $\nu_h[t]$ is to set $\nu_h[t] = 1/t$, $\forall h = 0,1,\dots, t, \; \forall t $. The accumulation point of the sequence of primal values $\{\hat{r}_{ij}^{u,l}[t]\}$ and $\{\hat{x}_{ij}^{l}[t]\}$ generated via \eqref{eq:primalrecovery1} and \eqref{eq:primalrecovery2} is then a feasible and optimal solution to the optimization problem in \eqref{eq:optimization2} \cite{Sherali96}. For this particular choice of convex combination weights, the sequences of primal values generated via \eqref{eq:primalrecovery1} and \eqref{eq:primalrecovery2} can be calculated recursively as
\begin{subequations}
	\begin{align}
		\hat{r}_{ij}^{u,l}[t] = \frac{t-1}{t}\hat{r}_{ij}^{u,l}[t-1] + \frac{1}{t}r_{ij}^{u,l}[t] \label{eq:recursive1}\\
		 \hat{x}_{ij}^{l}[t] = \frac{t-1}{t}\hat{x}_{ij}^{l}[t-1] + \frac{1}{t}x_{ij}^{l}[t] \label{eq:recursive2}
	\end{align}
\end{subequations}


\section{Protocol design }
\label{sec:implementation}

Since both the server and the intermediate nodes perform network coding operations on the data packets, the standard NDN protocol requires mechanisms that would permit to handle Interest messages and Data objects in a network coding enabled setting. In this section, we discuss the new features and functionalities that we introduce in the NDN architecture, in order to enable the processing of the Data objects and Interest messages when network coding is performed in the NDN nodes. We also present our content aware forwarding strategy that builds upon our rate allocation algorithm described in Section \ref{sec:algorithm} and aims at the efficient delivery of the video content to the clients at the best possible quality. Central to our protocol and forwarding strategy design is the requirement to maintain a sufficiently high innovative rate of network coded packets in order to ensure the timely delivery of the video content.

\subsection{Naming}
\label{sec:naming}

The naming scheme of our network coding enabled NDN architecture follows the standard hierarchical naming of the NDN protocol and has the general format \textit{/component1/}$\dots$\textit{/componentN/NCFlag/PacketId/GenIndx}. The \emph{NCFlag} component enables the identification and subsequent processing of Interest messages that express interest for network coded data.  The \emph{NCFlag} is a binary flag. The 0 value signifies a request for an original source packet, while the 1 value corresponds to a request for a network coded Data object. Thus, the \emph{NCFlag} permits the nodes to distinguish between the two types of Interest messages and to invoke the appropriate processing functions. The \emph{NCFlag} is followed by the \emph{PacketId} name component, which specifies the requested packet. When the \emph{NCFlag} is 0, the \emph{PacketId} is interpreted as the sequence number of the packet within the generation. The sequence number along with the generation index uniquely identify the packet within the entire sequence of source video packets. When the request is for a network coded Data object, the \emph{PacketId} is understood as the class which the network coded Data object belongs to. In this case, the name in the Interest message no longer specifies a unique Data object but rather refers to any network coded packet that belongs to this class and generation. The last new component of the naming structure is the \emph{GenIndx}. It encodes the index of the generation which the packet belongs to. 

\subsection{Bloom filter based forwarding}
\label{sec:bloomfilter}

Unlike the original NDN protocol, where the content name in the Interest message identifies a unique Data object, in our network coding enabled NDN architecture the Interest messages express a request for a network coded packet of a certain class and generation. In this case, the content name does not specify a unique packet, but rather a group of packets with similar information content. On the one hand, the random linear coding of data packets in the network nodes increases the content diversity in the network and facilitates the optimization of the forwarding strategy, since an Interest message requesting a network coded Data object of a certain class and generation can be consumed by any available linear combination of packets of this class and generation. This information diversity significantly contributes to the efficient use of bandwidth and caching resources. On the other hand, from the client's perspective, this approach introduces some degree of ambiguity since clients issue multiple Interest messages with the same content name in order to obtain all the packets of a certain class and generation. This impedes the use of pipelining and entails a potential risk of consuming two distinct Interest messages with the same linear combination of source packets, thus reducing the innovative packet rate. It is thus necessary to provide the NDN nodes with a mechanism that would permit to distinguish between two Interest messages with the same content name and to decide whether they can be consumed by the same Data object. 

In order to resolve the ambiguity created by the lack of unique mapping between content names and the data, we make use of Bloom filters \cite{Bloom70} in order to store some additional information about the Interest messages and the Data objects. Specifically, we add a new field that contains a Bloom filter in both the Interest message and the Data object. This Bloom filter is a compact representation of a set whose elements are the IDs of the clients. When a client generates an Interest message, it inserts its ID in the originally empty Bloom filter. As the Interest message is forwarded towards the location of the data, the content of its Bloom filter is modified by the forwarding strategy according to some rules derived based on the optimal rate allocation, as will be explained in Section \ref{sec:FWDstrategy}. The content of the Bloom filter included in an Interest message can be interpreted as the set of clients that are the destination nodes for the Data object that will consume this Interest message. Thus if, for example, two Interest messages with the same content name arrive at a node, they can be consumed by the same Data object as long as the intersection of their Bloom filters is empty, which means that the same data will not be forwarded multiple times to the same client. When a network coded Data object consumes an Interest message, the content of the Bloom filter of the Interest message is copied to the Bloom filter of the newly generated network coded Data object. In that way, the Data object contains a compact representation of the set of destination nodes where this Data can be potentially forwarded. Thus, when a Data object arrives at a node, it can only consume the Interest message whose Bloom filter contains a subset of the client IDs stored in the Bloom filter of the Data object. In addition, once this Data object has been forwarded to some of the clients whose IDs where originally in its Bloom filter, these clients are removed from the list of potential consumers of this data so as to avoid sending the same Data object to the same client multiple times. More details on the use of Bloom filters for the processing of Interest messages and Data objects will be given in Sections \ref{sec:IMprocessing} and \ref{sec:DOprocessing}, respectively.

\begin{figure}[t]
	\begin{center}
		\subfloat[]{\includegraphics[width = 0.42 \textwidth]{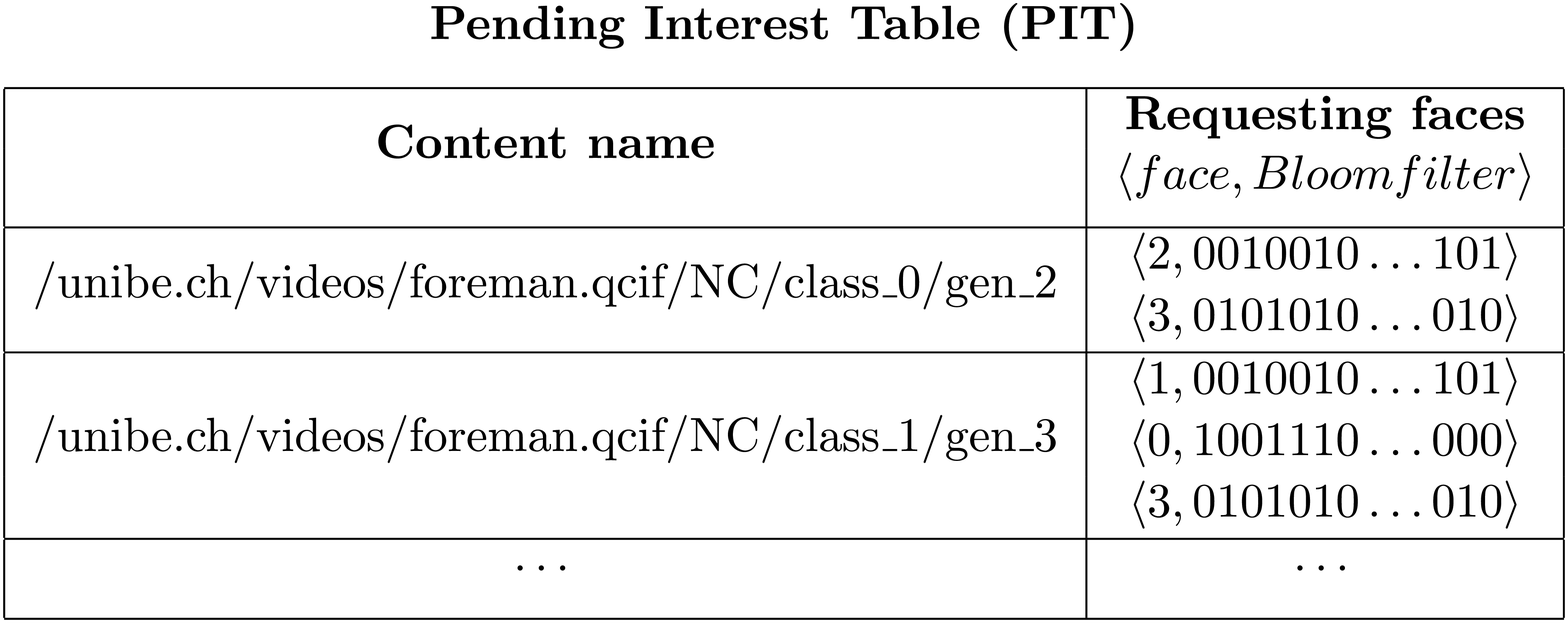}} \quad
		\subfloat[]{\includegraphics[width = 0.51 \textwidth]{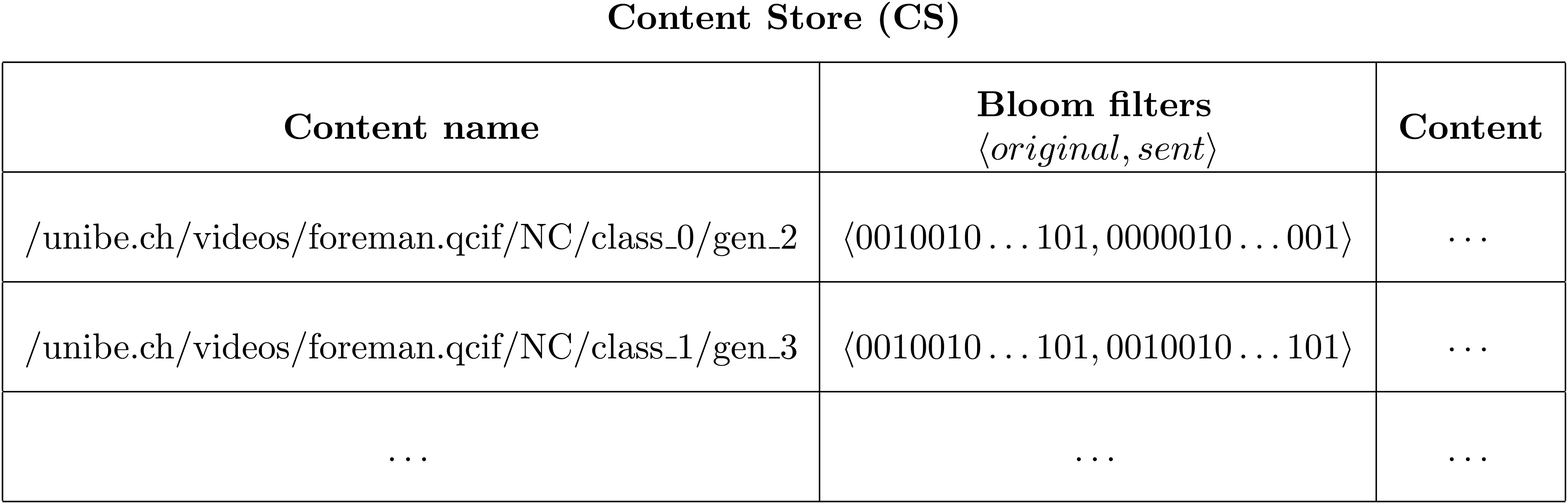}} 
	\end{center}
		\caption{Modified (a) Pending Interest Table (PIT) and (b) Content Store (CS) in our network coding enabled NDN architecture.}
	\label{fig:datastructures}
\end{figure}

Finally, since the Bloom filters are an essential element of our protocol design and play an important role in the handling of the packets, they are also present in both the Pending Interest Table (PIT) and the Content Store (CS). The modified tables are shown in Fig. \ref{fig:datastructures}. In the PIT, the list of incoming faces associated with some content name is replaced by a list of tuples $\langle face, Bloom \; filter\rangle$ , which indicates the incoming face and the Bloom filter of the Interest message which arrived on this face. Similarly, for every Data object that is stored in the CS, its Bloom filter is also stored in the CS along with the content name and the payload. Since the removal of elements from a Bloom filter is not permitted \cite{Bloom70}, a second Bloom filter is added for each data entry, which stores the ``sent'' information, \emph{i.e.}, the set of client IDs which the Data object has already been forwarded to.

\subsection{Handling of Interest messages}
\label{sec:IMprocessing}

\begin{algorithm}[t]
\baselineskip=17pt
\caption{PIT lookup procedure upon arrival of the Interest message $I$}
\label{algo:PITlookup}
\begin{algorithmic}[1]
\STATE \textbf{Input:} $I$, $\mathcal{P}_I$ 
\STATE \textbf{Output:} \emph{PendingInterestExists}, $I_x$
\STATE \textbf{Initialization:} \emph{PendingInterestExists} $\leftarrow$  \FALSE, \\$\mathcal{U}_I \leftarrow \{u \in \mathcal{U} \; | \; u \mbox{ in }\mbox{BF}_I\}$ 
\WHILE {$\mathcal{P}_I \neq \emptyset$}
\STATE Select the first $I^\prime$ in $\mathcal{P}_I$
\STATE Compute BF$_{I^\prime}^{union}$, \emph{i.e.} the union of Bloom filters in the list associated with $I^\prime$
\IF {$\mbox{BF}_{I^\prime}^{union} \cap \mbox{BF}_u \neq \mbox{BF}_u $ for all $u \in \mathcal{U}_I$}
\STATE \emph{PendingInterestExists} $\leftarrow $ \TRUE, $I_x \leftarrow I^\prime$,  $\mathcal{P}_I \leftarrow \emptyset$
\ELSE 
\STATE $\mathcal{P}_I \leftarrow \mathcal{P}_I \backslash I^\prime $
\ENDIF
\ENDWHILE
\RETURN \emph{PendingInterestExists}, $I_x$
\end{algorithmic}
\end{algorithm}

When an Interest message arrives on some face, the PIT is first checked for a pending Interest message requesting the same content. Unlike the standard NDN protocol, where a positive outcome is observed if the content name of the incoming Interest message matches the content name of some pending Interest message, this criterion is not sufficient for the PIT lookup procedure in the case of the network coded data. This insufficiency stems from the fact that the content name in our framework no longer specifies a unique Data object, but rather a collection of Data objects, \emph{i.e.}, the set of network coded packets that belong to the same class and generation. In order to obtain all the network coded packets of a certain class and generation, a client will issue as many Interest messages with the same content name as the number of packets in the specified class. It is therefore highly likely that two Interest messages with the same content name may be requesting two linearly independent network coded packets and cannot be consumed by the same Data object. In order to distinguish between the case where the incoming Interest message can be consumed by the same Data object as a pending Interest message and the case where the incoming Interest message must be treated as a request for a new Data object, additional criteria must be fulfilled for the PIT lookup procedure. These criteria are based upon the information that is stored in the Bloom filters of the Interest messages. 

Let $I$ denote the incoming Interest message. Algorithm \ref{algo:PITlookup} summarizes the PIT lookup for a pending Interest message upon the arrival of $I$. The algorithm takes as input the set $\mathcal{P}_I$ of pending Interest messages in the PIT whose content name matches the content name of $I$, and outputs the Boolean variable \emph{PendingInterestExists}, which indicates whether a matching pending Interest message was found. If a matching pending Interest message exists in the PIT, the algorithm also outputs a pointer $I_x$ to the PIT entry where it is stored. The PIT lookup proceeds as follows. For every pending Interest message $I^\prime$ in $\mathcal{P}_I$, we first compute the union BF$^{union}_{I^\prime}$ of the Bloom filters that are stored in the list of requesting faces of $I^\prime$. This union contains the set of clients whose request can be satisfied by the same Data object. We then compare the set of client IDs stored in BF$_{I^\prime}^{union}$ with the set $\mathcal{U}_I$ of the client IDs stored in the Bloom filter BF$_I$ of $I$. If none of the client IDs in $\mathcal{U}_I$ is present in BF$_{I^\prime}^{union}$, $I^\prime$ is considered as a matching pending Interest message. This essentially means that $I$ and $I^\prime$ can be consumed by the same Data object. It is worth mentioning that the use of the Bloom filters and the additional conditions on the PIT lookup procedure enable pipelining which would not be otherwise feasible due to the fact that the same content name describes more than one network coded Data objects. 


If a pending Interest message is found in the PIT, the arrival face of $I$ as well as its Bloom filter BF$_I$ are stored in the list of requesting faces of the entry pointed by $I_x$ and no further action is taken. If the PIT lookup procedure does not yield any matching pending Interest message, the Forwarding Information Base (FIB) is checked for a matching entry. The FIB lookup procedure is identical to the one in the standard NDN protocol and is based only on the content name. If a matching FIB entry is found, the Interest message is forwarded according to the forwarding strategy that will be described in Section \ref{sec:FWDstrategy}, and then inserted as a new entry in the PIT. 

If both the PIT and the FIB lookup procedures fail to produce a positive outcome, the CS is searched for a Data object that can potentially consume $I$. In order to determine whether the CS contains such a Data object, we once again make use of the information stored in the Bloom filters of $I$ and the Data objects in the CS. The CS lookup procedure upon the arrival of $I$ is summarized in Algorithm \ref{algo:CSlookup}. The algorithm takes as input the set $\mathcal{D}_I$ of Data objects in the CS, whose content name matches the content name of $I$. For every client $u$ whose ID is stored in the Bloom filter BF$_I$ of $I$, the algorithm searches the set $\mathcal{D}_I$ for a Data object $D$, that has the same client ID stored in its Bloom filter BF$_D$, but not in the Bloom filter BF$_D^{sent}$, which keeps track of the clients to whom $D$ has been already forwarded. If a Data object satisfying these criteria is found for all the clients in $\mathcal{U}_I$, then we consider that a Data object that matches $I$ can be generated from the Data objects that are stored in the CS. The algorithm outputs the Boolean variable \emph{MatchingDataExists}, which indicates whether a Data object can be generated to consume $I$. In case of a positive outcome, the algorithm also outputs a set $\mathcal{D}_{I}^\prime$ of Data objects that must be combined with RLNC in order to consume $I$. This CS lookup procedure guaranties that a Data object that has already been sent in response to an Interest message originating from some client $u$, will not be retransmitted multiple times to $u$.

\begin{algorithm}[t]
\baselineskip=17pt
\caption{CS lookup procedure upon arrival of the Interest message $I$}
\label{algo:CSlookup}
\begin{algorithmic}[1]
\STATE \textbf{Input:} $I$, $\mathcal{D}_I$
\STATE \textbf{Output:} \emph{MatchingDataExists}, $\mathcal{D}_I^\prime$
\STATE \textbf{Initialization:} $\mathcal{U}_I^\prime \leftarrow \emptyset$, \; $\mathcal{D}_I^\prime \leftarrow \emptyset$,\\
 \emph{MatchingDataExists} $\leftarrow$ \FALSE, $\mathcal{U}_I \leftarrow \{u \in \mathcal{U}\;|\; u \in \mbox{BF}_I \}$
\WHILE {$\mathcal{U}_I \neq \emptyset$ \textbf{ and } $\mathcal{D}_I \neq \emptyset$}
\STATE Select the first $D$ in $\mathcal{D}_I$
\FOR {every $u \in \mathcal{U}_I$}
\IF {$u \in\mbox{BF}_D \; \textbf{and} \;  u \notin \mbox{BF}_D^{sent}$} 
\STATE $\mathcal{D}_I^\prime \leftarrow \mathcal{D}_I^\prime \cup D$, $\mathcal{U}_I^\prime \leftarrow \mathcal{U}_I^\prime \cup u$
\ENDIF
\ENDFOR
\STATE $\mathcal{D}_I \leftarrow \mathcal{D}_I \backslash D$, $\mathcal{U}_I \leftarrow \mathcal{U}_I \backslash \mathcal{U}_I^\prime$, $\mathcal{U}_I^\prime \leftarrow \emptyset$
\ENDWHILE
\IF {$\mathcal{U}_I == \emptyset$}
\STATE  \emph{MatchingDataExists} $\leftarrow$ \TRUE
\ENDIF
\RETURN \emph{MatchingDataExists}, $\mathcal{D}_I^\prime$
\end{algorithmic}
\end{algorithm}

\begin{algorithm}[t]
\baselineskip=17pt
\caption{CS update procedure after transmission of Data object that consumes the Interest message $I$}
\label{algo:CSupdate}
\begin{algorithmic}[1]
\STATE \textbf{Input:} $I$, $\mathcal{D}_I^\prime$ 
\STATE \textbf{Initialization:} $\mathcal{U}_I \leftarrow \{u \in \mathcal{U}\;|\; u \in \mbox{BF}_I \}$
\WHILE {$\mathcal{U}_I \neq \emptyset$}
\STATE Pick $u \in \mathcal{U}_I$
\STATE Find $D \in \mathcal{D}_I^\prime \mbox{ s.t. } u \in \mbox{ BF}_D \textbf{ and } u \notin  \mbox{ BF}_D^{sent}$
\STATE Insert $u$ in BF$_D^{sent}$
\STATE $\mathcal{U}_I \leftarrow \mathcal{U}_I \backslash u$
\ENDWHILE
\end{algorithmic}
\end{algorithm}

If the outcome of the Algorithm \ref{algo:CSlookup} is true, a network coded Data object is generated by combining the Data objects in the set $D_I$ with RLNC.\footnote{Note that this is equivalent to combining only the Data objects in $\mathcal{D}_I^\prime$.} Its Bloom filter is set equal to the Bloom filter of $I$. The network coded packet is then scheduled for transmission on the arrival face of $I$. The CS is updated according to the procedure described in Algorithm \ref{algo:CSupdate}. In particular, the IDs of the clients that are stored in the Bloom filter BF$_I$ of $I$, are inserted in the Data objects in the set $\mathcal{D}_I^\prime$ that where identified as not yet transmitted to those clients. This update of the CS entries ensures that the network coded Data objects stored in CS will not be sent multiple times to the same client.

If after examining all three data structures, \emph{i.e.}, PIT, FIB and CS, a match is not found, the incoming Interest message is inserted as a new entry in the PIT and remains there until a matching Data object arrives at the node. The absence of a match in all three data structures indicates that the node has already forwarded the necessary number of Interest messages in order to receive a sufficient amount of innovative content but not all the data has yet arrived at the node.

\noindent \emph{Remark}: Explain why a PIT match is preferred over FIB match and FIB match is preferred over a CS match.

\subsection{Handling of Data objects}
\label{sec:DOprocessing}


When a Data object arrives, the timestamp of the video packet is first compared to the current time. If according to the time stamp the packet has expired, the Data object is discarded, since it is no longer useful for the clients. 
%
%
%
If the incoming Data object has not expired, a PIT lookup is performed for a matching pending Interest message. A PIT match is determined based on not only the content name, as in the standard NDN protocol, but also on the information stored in the Bloom filters of the incoming Data object and of the pending Interest messages in the PIT. The PIT lookup procedure upon arrival of an innovative Data object $D$ is described in Algorithm \ref{algo:PITlookupData}. The algorithm takes as input the set $\mathcal{P}_D$ of pending Interest messages in the PIT whose content name matches the content name of $D$, and the set $\mathcal{D}_D$ of Data objects stored in the CS, whose content name matches the content name of $D$. For every pending Interest message $I^\prime$ in the set $\mathcal{P}_D$, the CS lookup procedure described in Algorithm \ref{algo:CSlookup} is invoked in order to determine whether the data stored in the CS is sufficient to consume $I^\prime$. If a set of necessary Data objects is found, Algorithm \ref{algo:PITlookupData} outputs the pending Interest message $I$ that can be consumed as well as the set $\mathcal{D}_D^\prime$ of Data objects in CS that are required to consume $I$. A network coded Data object is then generated by combining all the Data objects in $\mathcal{D}_D$ with RLNC and is scheduled for transmission on all the faces in the list of requesting faces of the pending Interest $I$. The pending Interest message $I$ is then removed from the PIT and the CS is updated using the procedure described in Algorithm \ref{algo:CSupdate} with input $I$ and $\mathcal{D}_D^\prime$. The above procedure is repeated until no matching pending Interest message can be found. 

\begin{algorithm}[t]
\baselineskip=17pt
\caption{PIT lookup procedure upon arrival of an innovative Data object $D$}
\label{algo:PITlookupData}
\begin{algorithmic}[1]
\STATE \textbf{Input:} $\mathcal{P}_D$, $\mathcal{D}_D$
\STATE \textbf{Output:} \emph{PendingInteretsExists}, $\mathcal{D}^\prime_{D}$ and $I$ 
\STATE \textbf{Initialization:} \emph{PendingInteretsExists} $\leftarrow$ \FALSE, $\mathcal{D}_{D}^\prime \leftarrow \emptyset$
\WHILE {$\mathcal{P}_D \neq \emptyset$}
\STATE Select the first $I^\prime$ in $\mathcal{P}_D$
\STATE $\mathcal{U}_{I^\prime} \leftarrow \{u \in \mathcal{U} \; | \; u \in \mbox{BF}_{I^\prime} \}$
\STATE Run Algorithm \ref{algo:CSlookup} with $I^\prime$, $\mathcal{D}_D$ as input \label{step1}
\STATE Let \emph{MatchingDataExists}, $\mathcal{D}_D^\prime$ be the output of step \ref{step1}
\IF {\emph{MatchingDataExists} == \TRUE}
\STATE \emph{PendingInteretsExists} $\leftarrow$ \TRUE, $I \leftarrow I^\prime$, $\mathcal{P}_D \leftarrow \emptyset$ 
\ELSE
\STATE $\mathcal{P}_D\leftarrow \mathcal{P}_D \backslash I^\prime$
\ENDIF
\ENDWHILE
\RETURN \emph{PendingInteretsExists}, $\mathcal{D}^\prime_{D}$ and $I$ 
\end{algorithmic} 
\end{algorithm}

\begin{figure}[t]
	\begin{center}
		\subfloat[]{\includegraphics[width = 0.45 \textwidth]{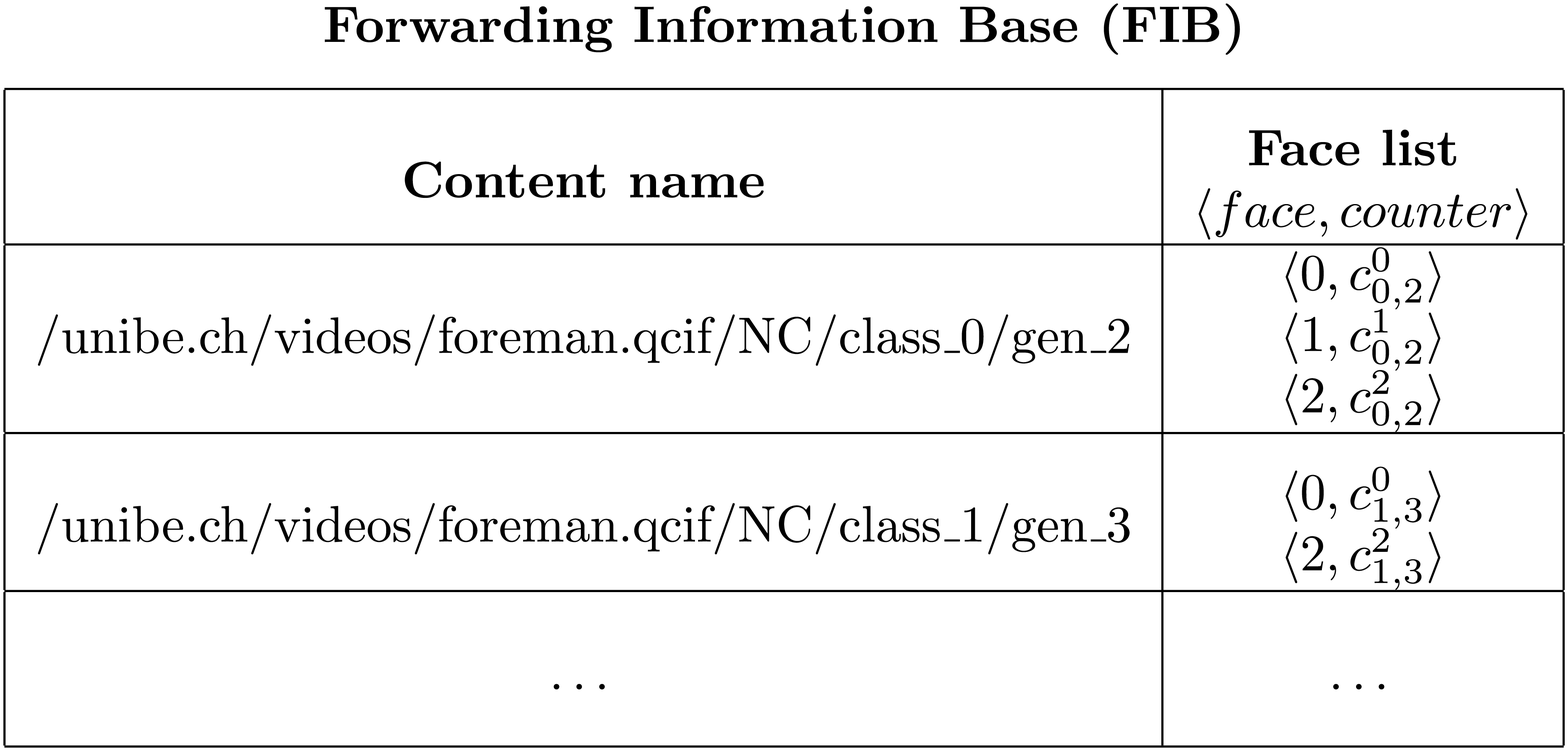}}
		\end{center}
		\caption{Modified Forwarding Information Base (FIB) in our network coding enabled NDN architecture.}
	\label{fig:FIB}
\end{figure}

\subsection{Forwarding strategy}
\label{sec:FWDstrategy}
The forwarding strategy controls the forwarding of Interest messages, when no matching pending Interest message can be found in the PIT, and performs the tasks of selecting the outgoing face and of modifying appropriately the Bloom filter of the forwarded Interest message in order to achieve the packet rate dictated by the rate allocation algorithm. 

The face selection mechanism is implemented in the Forwarding Information Base. The modified FIB table is illustrated in Fig. \ref{fig:FIB}. Each FIB entry consists of a content name, and a list of $\langle f,c_{l,g}^f \rangle$ tuples, where $f$ is the outgoing face and $c_{l,g}^f$ is a counter associated with the outgoing face $f$ and the content name. At the beginning of the streaming session, the FIB table is initialized with the entries that refer to all packet classes of the first $G$ generations. As the streaming session progresses, the entries that are related to the generations that have expired are deleted from the FIB, while new entries for more recent generations are inserted. This permits to keep the number of FIB entries low and at the same time to store all the information that is needed in order to manage requests within a given time window. For each new entry inserted in the FIB, the counters $c_{l,g}^f$ associated with the outgoing faces are initialized using the values obtained from the rate allocation algorithm. Specifically, we set $c_{l,g}^f = \tilde{z}_{ij}^{l}$ for every generation $g$, where $\tilde{z}_{ij}^{l}$ is the rate of the actual flow of Interest messages for network coded Data objects of class $l$ on the link $(i,j)$ and $f$ is the face associated with this link. When an FIB lookup is performed, the content name of the Interest message is compared against the content names in the entries of the FIB. If a FIB entry with a matching content name is found, the Interest message is forwarded on all the faces in the list for which the counter is non zero. Once the Interest message is scheduled for transmission, the corresponding counter is reduced by one. When all the counters in the list associated with some entry become zero, this entry is removed from the FIB and the node does not forward any more Interest messages with this content name.

The second task of the forwarding strategy is to insert the appropriate client IDs in the Bloom filter of the Interest message that is being forwarded. The selection procedure is described in Algorithm \ref{algo:BFselection}. The algorithm first computes the counter $p$ which indicates the number of Interest messages of class $l$ and generation $g$ that will have been transmitted on face $f$ including the Interest message that is being currently forwarded. Then, for every client, the algorithm examines whether this client's ID should be inserted in the Bloom filter of the Interest message. Specifically, the ID of client $u$ must be inserted in the Bloom filter of every $t$-th forwarded packet, where $t$ is computed in step \ref{step2}. To better illustrate the intuition behind this algorithm let us give an example. Let $\tilde{z}_{ij}^{l} = 9$, $\tilde{r}_{ij}^{u_1,l} = 9$ and $\tilde{r}_{ij}^{u_2,l} = 3$. Since the rate of the conceptual flow of client $u_1$ is equal to the actual rate of Interest messages, the ID of the client $u_1$ will be inserted in all Interest messages forwarded on this link. On the other hand, only 1/3 of the Interest messages originate from client $u_2$, which means that only 1/3 of the Data objects delivered on this link will be useful for client $u_2$. Therefore, the ID of the client $u_2$ will be inserted only in every third Interest message forwarded on this link.

\begin{algorithm}[t]
\baselineskip=17pt
\caption{Construction of the Bloom filter for an Interest message that is being forwarded}
\label{algo:BFselection}
\begin{algorithmic}[1]
\STATE \textbf{Input:} $c_{l,g}^f$, $\tilde{z}_{ij}^{l}$, $\tilde{r}_{ij}^{u,l}, \; \forall u \in \mathcal{U}$ 
\STATE \textbf{Output} BF${_I}$
\STATE \textbf{Initialization: } BF${_I} \leftarrow \emptyset$, $p \leftarrow \tilde{z}_{ij}^{l} - c_{l,g}^f + 1$
\FOR {every $ u \in \mathcal{U}$}
\IF {$\tilde{r}_{ij}^{u,l} \neq 0$}
\STATE $t \leftarrow \left \lfloor \frac{\tilde{z}_{ij}^{l}}{\tilde{r}_{ij}^{u,l}} \right \rfloor$ \label{step2}
\IF {$p \mod t = =0$ \textbf{and} ${p}/{t} \leq \tilde{r}_{ij}^{u,l} $}
\STATE Insert $u$ in BF$_I$
\ENDIF
\ENDIF
\ENDFOR
\RETURN BF$_I$
\end{algorithmic}
\end{algorithm}

\section{Performance evaluation}
\label{sec:results}

\subsection{Evaluation scenario}
\label{sec:simulationsetup}

we consider that the server encodes the Forman video sequence in CIF format into three quality layers with the SVC extension of the H.264 video compression standard \cite{SVC2005}. The GOP size is set to 30 frames and the frame rate is set to 30 fps. Each GOP is packetized into 78 packets, which consist of $a_0 = 38$ base layer packets, $a_1 = 15$ first enhancement layer packets and $a_2 = 20$ second enhancement layer packets. The size of the Data objects is $p_D = 1600$ bytes and the size of the Interest Messages is $p_I = 200$ bytes. The generation size is set equal to the size of the GOP, \emph{i.e.}, 78 packets. That means that each generation corresponds to one second of video. The duration of the entire video sequence is 40 seconds, \emph{i.e.}, there are 40 generations in total. The video quality achieved after decoding each of the three video layers is $q_0 = 36.48$ dB, $q_1 = 37.82$ dB and $q_2 = 39.09$ dB, respectively. The cost coefficients used in the rate allocation algorithm are $c_0 = 0.01$, $c_1 = 0.02 $ and $c_2 = 0.025$, and are chosen according to \eqref{eq:costconstraint1} and \eqref{eq:costconstraint2}. 

The size of the Galois field, where the network coding operations are performed, is chosen equal to $2^8$. This field size constitutes a common design choice for practical applications using RLNC, as it achieves a good trade-off between the probability of generating non-innovative packets due to the random selection of the network coding coefficients, and the data overhead due to the inclusion of the network coding coefficients in the header \cite{ChouAllerton2003}.

\begin{figure}
	\begin{center}
		\includegraphics[width = 0.4 \textwidth]{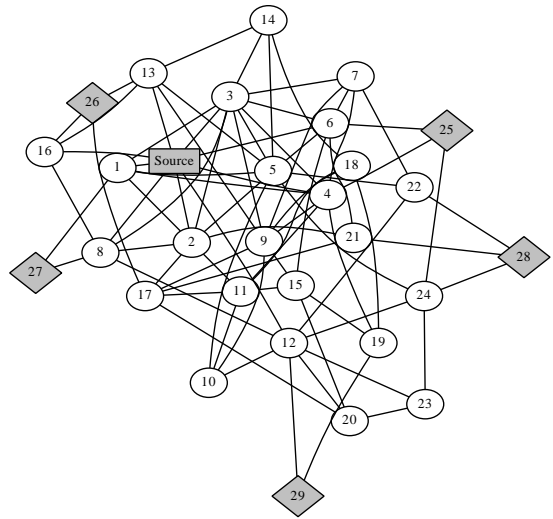}
	\end{center}
	\caption{PlanetLab topology consisting of one source node (node 0), 24 intermediate nodes and 5 client nodes (nodes 25 - 29.)}
\label{fig:planetlab_topology}
\end{figure}

We evaluate the performance of our proposed architecture on the network topology depicted in Fig. \ref{fig:planetlab_topology}. This topology was constructed based on real data collected by the PlanetLab project \cite{Planetlab2015}. The topology consists of one source node, 5 client nodes (nodes 25-29) and 24 helpers. To generate this topology we followed the procedure described in \cite{ClejuTMM2011}. 

\input{convergence_double}

\subsection{Convergence}
\label{sec:convergence}

We first investigate the convergence of our rate allocation algorithm presented in Section \ref{sec:rateallocation}. In Fig. \ref{fig:convergence} we illustrate the evolution of the cumulative rate of Interest messages sent by node 29 with the number of iterations of the rate allocation algorithm presented in Section \ref{sec:rateallocation} for different values of the link bandwidth. Each figure shows the convergence of the rate of Interest messages requesting network coded packets that belong to the three available packet classes. We can see that when the link bandwidth is low, the client requests only network coded packets from class 0 which consists of packets that belong to the base layer. The rate of Interest messages requesting class 0 packets converges to the value of 38 packets/sec which is the encoding rate of the base layer. As the link bandwidth increases, the client starts to request also class 1 network coded packets, which results in decoding the first enhancement layer. The rate of the Interest messages requesting class 1 packets converges to the value of the encoding rate of the first enhancement layer, which is equal to 15 packets/sec. Finally, for sufficiently large values of bandwidth, the client requests packets from all three classes which permits to decode the second enhancement layer. It is worth noting that among several possible optimal solutions, our content-aware rate allocation algorithm favors the solution that allocates the rate according to the importance of the delivered packets. In particular, we can observe that in cases where apart from the base layer additional enhancement layers can be decoded, each class of packets is requested at its encoding rate which minimizes the cost introduced in \eqref{eq:costfunction} and reflects the prioritization of the packet classes according to their importance.

\subsection{Video quality evaluation} 
\label{sec:simresults}

\begin{figure*}[t]
\centering
	\begin{tikzpicture}
		\begin{axis}[%
			width=0.42\textwidth,
			height=0.15\textwidth,
			at={(0\textwidth,0\textwidth)},
			scale only axis,
			xmin=281.25,
			xmax=562.5,
			xlabel={Link bandwidth (Kbps)},
			xmajorgrids,
			ymin=35,
			ymax=40,
			ylabel={PSNR (dB)},
			ymajorgrids,
			yticklabel style = {font=\footnotesize},
			xticklabel style = {font=\footnotesize},
			title style={font=\footnotesize},
			xlabel style={font=\footnotesize},
			ylabel style={font=\footnotesize},
			title={Node 25},
			legend style={at={(0.97,0.03)},anchor=south east,legend cell align=left,align=left,draw=white!15!black,font=\footnotesize}
			]
			\addplot [color=blue,solid,mark=diamond,mark options={solid}]
				table[row sep=crcr]{%
				281.25		37.82 \\
				323.4375	37.82 \\
				351.5625	39.09 \\	
				365.625		39.09 \\
				379.6875	39.09 \\
				421.875		39.09 \\
				478.125		39.09 \\
				520.3125	39.09 \\
				562.5		39.09 \\
			};
			\addlegendentry{UB};

			\addplot [color=red,solid,mark=asterisk,mark options={solid}]
				table[row sep=crcr]{%
				281.25		37.82 \\
				323.4375	37.82 \\
				351.5625	39.09 \\	
				365.625		39.09 \\
				379.6875	39.09 \\
				421.875		39.09 \\
				478.125		39.09 \\
				520.3125	39.09 \\
				562.5		39.09 \\
			};
			\addlegendentry{EXP};
			
			\addplot [color=green,solid,mark=triangle,mark options={solid}]
				table[row sep=crcr]{%
				281.25		37.4331 \\                   
				323.4375	37.5922 \\
				351.5625	38.8791 \\	
				365.625		38.7271 \\
				379.6875	38.8568 \\
				421.875		38.8073 \\
				478.125		38.7919 \\
				520.3125	38.6881 \\    
				562.5		38.6380 \\
			};
			\addlegendentry{SIM};
			
		\end{axis}
	\end{tikzpicture}%
	~
	\begin{tikzpicture}
		\begin{axis}[%
			width=0.42\textwidth,
			height=0.15\textwidth,
			at={(0\textwidth,0\textwidth)},
			scale only axis,
			xmin=281.25,
			xmax=562.5,
			xlabel={Link bandwidth (Kbps)},
			xmajorgrids,
			ymin=35,
			ymax=40,
			ylabel={PSNR (dB)},
			ymajorgrids,
			yticklabel style = {font=\footnotesize},
			xticklabel style = {font=\footnotesize},
			xlabel style={font=\footnotesize},
			ylabel style={font=\footnotesize},
			title style={font=\footnotesize},
			title={Node 26},
			legend style={at={(0.97,0.03)},anchor=south east,legend cell align=left,align=left,draw=white!15!black,font=\footnotesize}
			]
			\addplot [color=blue,solid,mark=diamond,mark options={solid}]
				table[row sep=crcr]{%
				281.25		37.82 \\ 
				323.4375	37.82 \\
				351.5625	39.09 \\	
				365.625		39.09 \\
				379.6875	39.09 \\
				421.875		39.09 \\
				478.125		39.09 \\
				520.3125	39.09 \\
				562.5		39.09 \\
			};
			\addlegendentry{UB};

			\addplot [color=red,solid,mark=asterisk,mark options={solid}]
				table[row sep=crcr]{%
				281.25		37.82 \\
				323.4375	37.82 \\
				351.5625	39.09 \\	
				365.625		39.09 \\
				379.6875	39.09 \\
				421.875		39.09 \\
				478.125		39.09 \\
				520.3125	39.09 \\
				562.5		39.09 \\
			};
			\addlegendentry{EXP};
			
			\addplot [color=green,solid,mark=triangle,mark options={solid}]
				table[row sep=crcr]{%
				281.25		37.5566 \\                       
				323.4375	37.6472 \\
				351.5625	38.6650 \\	
				365.625		38.8438 \\
				379.6875	38.9167 \\
				421.875		38.9641 \\
				478.125		39.0029 \\
				520.3125	38.9125  \\    
				562.5		38.5413 \\
			};
			\addlegendentry{SIM};
			
		\end{axis}
	\end{tikzpicture}%
	
	\begin{tikzpicture}
		\begin{axis}[%
			width=0.42\textwidth,
			height=0.15\textwidth,
			at={(0\textwidth,0\textwidth)},
			scale only axis,
			xmin=281.25,
			xmax=562.5,
			xlabel={Link bandwidth [Kbps]},
			xmajorgrids,
			ymin=35,
			ymax=40,
			ylabel={PSNR (dB)},
			ymajorgrids,
			yticklabel style = {font=\footnotesize,xshift=0.5ex},
			xticklabel style = {font=\footnotesize,yshift=0.5ex},
			xlabel style={font=\footnotesize},
			ylabel style={font=\footnotesize},
			title style={font=\footnotesize},
			title={Node 27},
			legend style={at={(0.97,0.03)},anchor=south east,legend cell align=left,align=left,draw=white!15!black, font=\footnotesize}
			]
			\addplot [color=blue,solid,mark=diamond,mark options={solid}]
				table[row sep=crcr]{%
				281.25		36.48 \\
				323.4375	36.48 \\
				351.5625	36.48 \\	
				365.625		36.48 \\
				379.6875	37.82 \\
				421.875		37.82 \\
				478.125		37.82 \\
				520.3125	39.09 \\
				562.5		39.09 \\	
			};
			\addlegendentry{UB};

			\addplot [color=red,solid,mark=asterisk,mark options={solid}]
				table[row sep=crcr]{%
				281.25		36.48 \\ 
				323.4375	36.48 \\
				351.5625	36.48 \\	
				365.625		36.48 \\
				379.6875	37.82 \\
				421.875		37.82 \\
				478.125		37.82 \\
				520.3125	39.09 \\
				562.5		39.09 \\
			};
			\addlegendentry{EXP};
			
			\addplot [color=green,solid,mark=triangle,mark options={solid}]
				table[row sep=crcr]{%
				281.25		36.2974 \\                       
				323.4375	36.3941 \\
				351.5625	36.3457 \\	
				365.625		36.3887 \\
				379.6875	37.7170 \\
				421.875		37.4032 \\
				478.125		37.6220 \\
				520.3125	38.9032 \\    
				562.5		38.6185 \\
			};
			\addlegendentry{SIM};
			
		\end{axis}
	\end{tikzpicture}%
	~
	\begin{tikzpicture}
		\begin{axis}[%
			width=0.42\textwidth,
			height=0.15\textwidth,
			at={(0\textwidth,0\textwidth)},
			scale only axis,
			xmin=281.25,
			xmax=562.5,
			xlabel={Link bandwidth (Kbps)},
			xmajorgrids,
			ymin=35,
			ymax=40,
			ylabel={PSNR (dB)},
			ymajorgrids,
			yticklabel style = {font=\footnotesize,xshift=0.5ex},
			xticklabel style = {font=\footnotesize,yshift=0.5ex},
			xlabel style={font=\footnotesize},
			ylabel style={font=\footnotesize},
			title style={font=\footnotesize},
			title={Node 28},
			legend style={at={(0.97,0.03)},anchor=south east,legend cell align=left,align=left,draw=white!15!black,font=\footnotesize}
			]
			\addplot [color=blue,solid,mark=diamond,mark options={solid}]
				table[row sep=crcr]{%
				281.25		37.82 \\
				323.4375	37.82 \\
				351.5625	39.09 \\	
				365.625		39.09 \\
				379.6875	39.09 \\
				421.875		39.09 \\
				478.125		39.09 \\
				520.3125	39.09 \\
				562.5		39.09 \\	
			};
			\addlegendentry{UB};

			\addplot [color=red,solid,mark=asterisk,mark options={solid}]
				table[row sep=crcr]{%
				281.25		37.82 \\ 
				323.4375	37.82 \\
				351.5625	39.09 \\	
				365.625		39.09 \\
				379.6875	39.09 \\
				421.875		39.09 \\
				478.125		39.09 \\
				520.3125	39.09 \\
				562.5		39.09 \\
			};
			\addlegendentry{EXP};
			
			\addplot [color=green,solid,mark=triangle,mark options={solid}]
				table[row sep=crcr]{%
				281.25		37.5529 \\                       
				323.4375	37.3415 \\
				351.5625	38.7044 \\	
				365.625		38.6374 \\
				379.6875	38.8946 \\
				421.875		38.9431 \\
				478.125		38.8455 \\
				520.3125	39.0181 \\    
				562.5		38.7219 \\
			};
			\addlegendentry{SIM};
			
		\end{axis}
	\end{tikzpicture}%
	
	\begin{tikzpicture}
		\begin{axis}[%
			width=0.42\textwidth,
			height=0.15\textwidth,
			at={(0\textwidth,0\textwidth)},
			scale only axis,
			xmin=281.25,
			xmax=562.5,
			xlabel={Link bandwidth (Kbps)},
			xmajorgrids,
			ymin=35,
			ymax=40,
			ylabel={PSNR (dB)},
			ymajorgrids,
			yticklabel style = {font=\footnotesize,xshift=0.5ex},
			xticklabel style = {font=\footnotesize,yshift=0.5ex},
			xlabel style={font=\footnotesize},
			ylabel style={font=\footnotesize},
			title style={font=\footnotesize},
			title={Node 29},
			legend style={at={(0.97,0.03)},anchor=south east,legend cell align=left,align=left,draw=white!15!black, font=\footnotesize}
			]
			\addplot [color=blue,solid,mark=diamond,mark options={solid}]
				table[row sep=crcr]{%
				281.25		36.48 \\
				323.4375	36.48 \\
				351.5625	36.48 \\	
				365.625		36.48 \\
				379.6875	37.82 \\
				421.875		37.82 \\
				478.125		37.82 \\
				520.3125	39.09 \\
				562.5		39.09 \\	
			};
			\addlegendentry{UB};

			\addplot [color=red,solid,mark=asterisk,mark options={solid}]
				table[row sep=crcr]{%
				281.25		36.48 \\
				323.4375	36.48 \\
				351.5625	36.48 \\	
				365.625		36.48 \\
				379.6875	37.82 \\
				421.875		37.82 \\
				478.125		37.82 \\
				520.3125	39.09 \\
				562.5		39.09 \\
			};
			\addlegendentry{EXP};
			
			\addplot [color=green,solid,mark=triangle,mark options={solid}]
				table[row sep=crcr]{%
				281.25		36.3135 \\                       
				323.4375	36.1524 \\
				351.5625	36.3189 \\	
				365.625		36.3243 \\
				379.6875	37.7120 \\
				421.875		37.5260 \\
				478.125		37.6381 \\
				520.3125	38.9447 \\    
				562.5		38.5677 \\
			};
			\addlegendentry{SIM};
			
		\end{axis}
	\end{tikzpicture}%
	
	\caption{Average PSNR of the decoded video as a function of the links' bandwidth.}
	\label{fig:quality_vs_bwd}
\end{figure*}
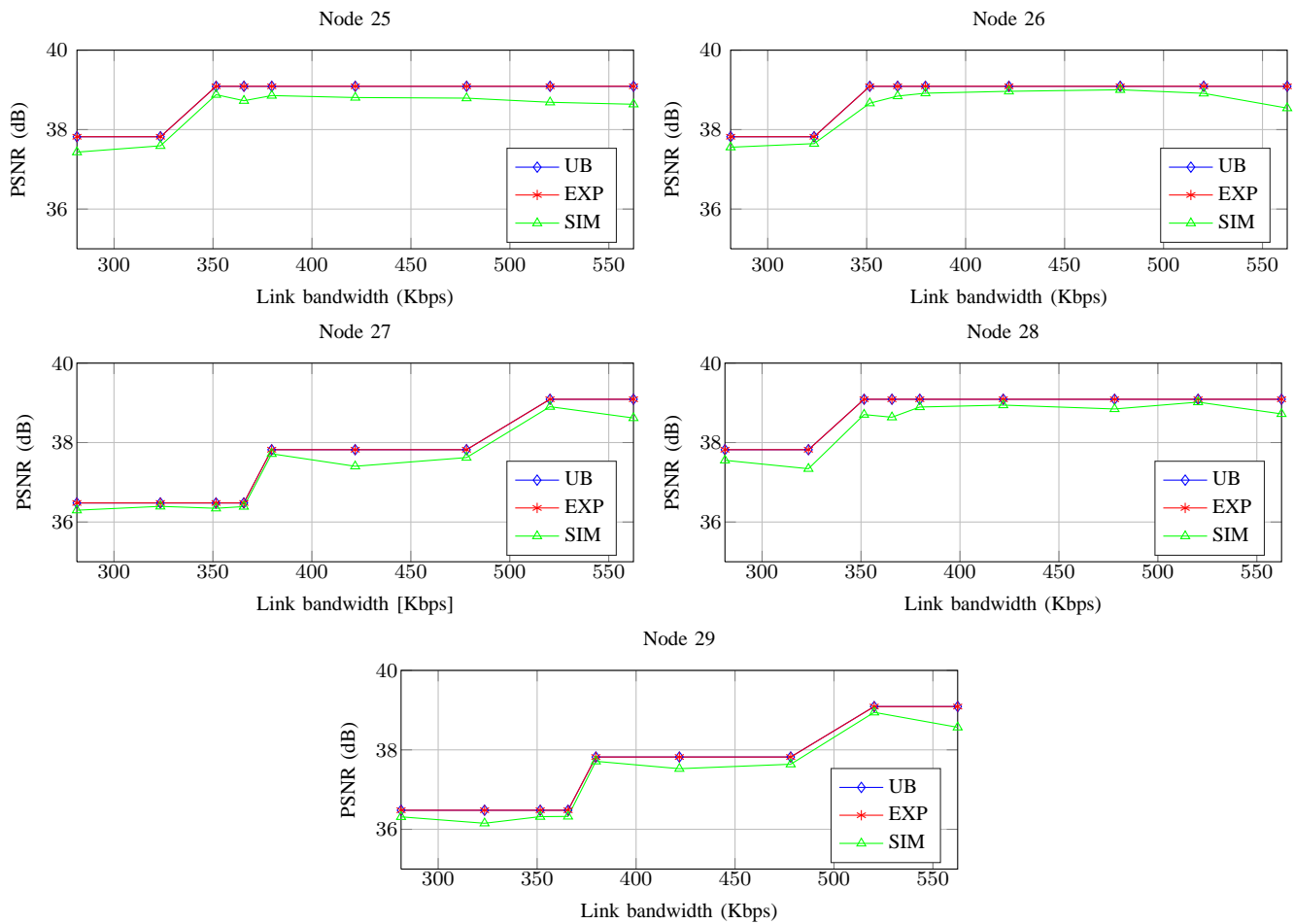

In order to validate our architecture, we have implemented a customized simulator in NS-3 \cite{ns3} with the main components of the NDN architecture as well as the additional features described in Section \ref{sec:implementation}, that enable the delivery of network coded scalable content. Each point in the figures is the average of 100 simulations. We assume that the clients join the streaming session with a random delay which is much smaller than the playback delay in order to create some randomness in the order in which Interest messages arrive at intermediate nodes. Another source of randomness is the random selection of coding coefficients during the network coding operations.

We evaluate the proposed protocol for the delivery of scalable video in terms of quality. In Fig. \ref{fig:quality_vs_bwd} we present the evolution of the PSNR of the delivered video with respect to the links' bandwidth for each network client. UB denotes the upper bound on the video quality that can be attained at the client. The upper bound is calculated based on the maximum achievable flow between each client and the source assuming that the client is using all the network resources. EXP stands for the expected value of the PSNR based on the rate allocation solution obtained with our content-aware rate allocation algorithm. Finally, SIM denotes the PSNR values obtained from the simulation of our network coding enabled NDN protocol equipped with the forwarding strategy described in Section \ref{sec:FWDstrategy}. We can see that the performance of our proposed forwarding strategy is very close to the expected one with a deviation that does not exceed 0.5 dB. Moreover, we note that as the links' bandwidth increases, the clients are able to decode more video layers and improve the quality of the delivered video. This improvement is attributed to the use of PRLNC that allows to optimally use the available resources. Our forwarding strategy exploits the optimal rate allocation in order to decide which Interest messages to forward on which faces. This ensures that there are no packet replicas and the rate of redundant packets is minimized. The small degradation in the quality of the delivered video compared to the optimal performance is caused by the randomness in the selection of the network coding coefficients which may result in the generation of linearly dependent network coded packets. Since the decoding of the $l$-th class of network coded packets requires that all the lower packet classes can be decoded, the delivery of even a single non-innovative packet of class $k$ causes a failure in decoding all classes higher or equal to $k$. 

\begin{figure*}[t]
\centering
	\begin{tikzpicture}
		\begin{axis}[%
			width=0.25\textwidth,
			height=0.15\textwidth,
			at={(0\textwidth,0\textwidth)},
			scale only axis,
			xmin=0,
			xmax=40,
			xlabel={Time (sec)},
			xmajorgrids,
			ymin=30,
			ymax=40,
			ylabel={PSNR (dB)},
			ymajorgrids,
			yticklabel style = {font=\footnotesize},
			xticklabel style = {font=\footnotesize},
			title style={font=\footnotesize,yshift=-1ex},
			title={Link bandwdith: 330 Kbps},
			xlabel style = {font=\footnotesize},
			ylabel style = {font=\footnotesize},
			legend style={at={(0.03,0.03)},anchor=south west,legend cell align=left,align=left,draw=white!15!black, font=\footnotesize}
			]
			\addplot [color=blue,solid,line width = 1.8pt]
				table[row sep=crcr]{%
				1	37.5784 \\
				2	37.8066 \\
				3	37.8066 \\
				4	37.3636 \\
				5	37.5918 \\
				6	37.7798 \\
				7	37.82 \\
				8	37.3636 \\
				9	37.565 \\
				10	37.3636 \\
				11	37.5918 \\
				12	37.5784 \\
				13	37.7932 \\
				14	37.82 \\
				15	37.565 \\
				16	37.3502 \\
				17	37.7798 \\
				18	37.7664 \\
				19	37.5784 \\
				20	37.7932 \\
				21	37.5918 \\ 
				22	37.82 \\
				23	37.3636 \\ 
				24	37.7798 \\
				25	37.8066 \\
				26	37.7932 \\
				27	37.8066 \\
				28	37.122 \\
				29	37.82 \\
				30	37.8066 \\
				31	37.5918 \\
				32	37.82 \\
				33	37.5784 \\
				34	37.82 \\
				35	37.82 \\
				36	37.5784 \\
				37	37.5784 \\
				38	37.3636	 \\
				39	37.7932 \\
				40	37.5784 \\
			};
			\addlegendentry{SIM - Node 26};

			\addplot [color=red,dashdotted,line width = 1.5pt]
				table[row sep=crcr]{%
			1.0000   36.0504 \\
    2.0000   36.4800\\
    3.0000   36.2652\\
    4.0000   36.4800\\
    5.0000   36.4800\\
    6.0000   36.2652\\
    7.0000   36.4800\\
    8.0000   36.4800\\
    9.0000   36.4800\\
   10.0000   36.4800\\
   11.0000   36.4800\\
   12.0000   36.4800\\
   13.0000   36.4800\\
   14.0000   36.4800\\
   15.0000   36.4800\\
   16.0000   36.2652\\
   17.0000   36.4800\\
   18.0000   36.4800\\
   19.0000   36.2652 \\
   20.0000   36.2652\\
   21.0000   36.4800\\
   22.0000   36.4800\\
   23.0000   36.2652\\
   24.0000   36.2652\\
   25.0000   36.4800\\
   26.0000   36.2652\\
   27.0000   36.4800\\
   28.0000   36.0504\\
   29.0000   36.4800\\
   30.0000   36.0504\\
   31.0000   36.4800\\
   32.0000   36.4800\\
   33.0000   36.4800\\
   34.0000   36.2652\\
   35.0000   36.2652\\
   36.0000   36.4800\\
   37.0000   36.4800\\
   38.0000   36.4800\\
   39.0000   36.4800\\
   40.0000   36.4800\\
			};
			\addlegendentry{SIM - Node 27};
		\end{axis}
	\end{tikzpicture}%
		\begin{tikzpicture}
		\begin{axis}[%
			width=0.25\textwidth,
			height=0.15\textwidth,
			at={(0\textwidth,0\textwidth)},
			scale only axis,
			xmin=0,
			xmax=40,
			xlabel={Time (sec)},
			xmajorgrids,
			ymin=30,
			ymax=40,
			ylabel={PSNR (dB)},
			ymajorgrids,
			yticklabel style = {font=\footnotesize},
			xticklabel style = {font=\footnotesize},
			title style={font=\footnotesize,yshift=-1ex},
			title={Link bandwidth: 480 Kbps},
			xlabel style = {font=\footnotesize},
			ylabel style = {font=\footnotesize},
			legend style={at={(0.03,0.03)},anchor=south west,legend cell align=left,align=left,draw=white!15!black, font=\footnotesize}
			]
			\addplot [color=blue,solid,line width = 1.8pt]
				table[row sep=crcr]{%
				  1.0000   38.8491 \\
    2.0000   39.0378 \\
    3.0000   39.0519 \\
    4.0000   38.9863 \\
    5.0000   38.9609 \\
    6.0000   39.0773 \\
    7.0000   38.8230 \\
    8.0000   39.0639 \\
    9.0000   39.0512 \\
   10.0000   39.0773 \\
   11.0000   38.8491 \\
   12.0000   39.0773 \\
   13.0000   39.0512 \\
   14.0000   39.0385 \\
   15.0000   38.8110 \\
   16.0000   38.7969 \\
   17.0000   39.0519 \\
   18.0000   38.8103 \\
   19.0000   39.0900 \\
   20.0000   39.0773 \\
   21.0000   38.8237 \\
   22.0000   39.0639 \\
   23.0000   39.0900 \\
   24.0000   39.0378 \\
   25.0000   39.0773 \\
   26.0000   38.8237 \\
   27.0000   39.0378 \\
   28.0000   39.0378 \\
   29.0000   39.0519 \\
   30.0000   39.0639 \\
   31.0000   39.0512 \\
   32.0000   39.0773 \\
   33.0000   39.0519 \\
   34.0000   39.0639 \\
   35.0000   39.0900 \\
   36.0000   39.0639 \\
   37.0000   39.0378 \\
   38.0000   39.0646 \\
   39.0000   38.8103 \\
   40.0000   39.0639 \\
			}; 
			\addlegendentry{SIM - Node 26};

			\addplot [color=red,dashdotted,line width = 1.5pt]
				table[row sep=crcr]{%
			1.0000   36.8938\\
    2.0000   37.7932 \\
    3.0000   37.3502 \\
    4.0000   37.5784 \\
    5.0000   37.5650 \\
    6.0000   37.7798 \\
    7.0000   37.5784 \\
    8.0000   37.5784 \\
    9.0000   37.5650 \\
   10.0000   37.7530 \\
   11.0000   37.5650 \\
   12.0000   37.5918 \\
   13.0000   37.8200 \\
   14.0000   37.3502 \\
   15.0000   37.5918 \\
   16.0000   37.7932 \\
   17.0000   37.5650 \\
   18.0000   37.8066 \\
   19.0000   37.5382 \\
   20.0000   37.7932 \\
   21.0000   37.5784 \\
   22.0000   37.7798 \\
   23.0000   37.3636 \\ 
   24.0000   37.7664 \\
   25.0000   37.5650 \\
   26.0000   37.5784 \\
   27.0000   37.5784 \\
   28.0000   37.8200 \\
   29.0000   37.7932 \\
   30.0000   37.5650 \\
   31.0000   37.7932 \\ 
   32.0000   37.7932 \\
   33.0000   37.3368 \\
   34.0000   37.8066 \\
   35.0000   37.5650 \\
   36.0000   37.5784 \\
   37.0000   37.5650 \\
   38.0000   37.8200 \\
   39.0000   37.5784 \\
   40.0000   37.8066 \\
			};
			\addlegendentry{SIM - Node 27};
		\end{axis}
	\end{tikzpicture}%
	\begin{tikzpicture}
		\begin{axis}[%
			width=0.25\textwidth,
			height=0.15\textwidth,
			at={(0\textwidth,0\textwidth)},
			scale only axis,
			xmin=0,
			xmax=40,
			xlabel={Time (sec)},
			xmajorgrids,
			ymin=30,
			ymax=40,
			ylabel={PSNR (dB)},
			ymajorgrids,
			yticklabel style = {font=\footnotesize},
			xticklabel style = {font=\footnotesize},
			title style={font=\footnotesize,yshift=-1ex},
			title={Link bandwidth: 560 Kbps},
			xlabel style = {font=\footnotesize},
			ylabel style = {font=\footnotesize},
			legend style={at={(0.03,0.03)},anchor=south west,legend cell align=left,align=left,draw=white!15!black, font=\footnotesize}
			]
			\addplot [color=blue,solid,line width = 1.8pt]
				table[row sep=crcr]{%
				  1.0000   39.0773 \\
    2.0000   39.0378 \\
    3.0000   38.7715 \\
    4.0000   38.8491 \\
    5.0000   38.8364 \\
    6.0000   39.0646 \\
    7.0000   39.0773 \\
    8.0000   39.0385 \\
    9.0000   38.7976 \\
   10.0000   39.0900 \\
   11.0000   39.0900 \\
   12.0000   39.0646 \\
   13.0000   39.0900 \\
   14.0000   39.0646 \\
   15.0000   39.0519 \\
   16.0000   38.8491 \\
   17.0000   39.0646 \\
   18.0000   39.0900 \\
   19.0000   39.0519 \\
   20.0000   38.8491 \\
   21.0000   38.5694 \\
   22.0000   38.8491 \\
   23.0000   38.8103 \\
   24.0000   38.3546 \\
   25.0000   39.0900 \\
   26.0000   39.0639 \\
   27.0000   38.5955 \\
   28.0000   39.0519 \\
   29.0000   39.0392 \\
   30.0000   39.0519 \\
   31.0000   38.3673 \\
   32.0000   39.0385 \\
   33.0000   38.5821 \\
   34.0000   38.8237 \\
   35.0000   38.8364 \\
   36.0000   38.7461 \\
   37.0000   39.0251 \\
   38.0000   38.7969 \\
   39.0000   38.8110 \\
   40.0000   39.0900 \\
			}; 
			\addlegendentry{SIM - Node 26};

			\addplot [color=red,dashdotted,line width = 1.5pt]
				table[row sep=crcr]{%
			1.0000   38.7969 \\
    2.0000   39.0646 \\
    3.0000   39.0385 \\
    4.0000   39.0900 \\
    5.0000   38.7969 \\
    6.0000   38.3412 \\
    7.0000   38.6082 \\
    8.0000   38.8237 \\
    9.0000   39.0900 \\
   10.0000   39.0258 \\
   11.0000   38.8364 \\
   12.0000   39.0512 \\
   13.0000   38.8230 \\
   14.0000   39.0900 \\
   15.0000   39.0900 \\
   16.0000   39.0512 \\
   17.0000   38.8491 \\
   18.0000   38.6082 \\
   19.0000   38.8491 \\
   20.0000   38.7581 \\
   21.0000   39.0900 \\
   22.0000   38.7581 \\
   23.0000   38.8237 \\
   24.0000   39.0639 \\
   25.0000   38.8491 \\
   26.0000   38.8237 \\
   27.0000   38.8103 \\
   28.0000   39.0900 \\
   29.0000   39.0251 \\
   30.0000   38.5955 \\
   31.0000   39.0639 \\
   32.0000   39.0646 \\
   33.0000   38.8491 \\
   34.0000   39.0773 \\
   35.0000   39.0639 \\
   36.0000   38.8364 \\
   37.0000   38.8364 \\
   38.0000   39.0639 \\
   39.0000   38.8364 \\
   40.0000   38.8230 \\
			};
			\addlegendentry{SIM - Node 27};
		\end{axis}
	\end{tikzpicture}%
	\caption{Average PSNR of the decoded video versus time.}
	\label{fig:quality_vs_time}
\end{figure*}
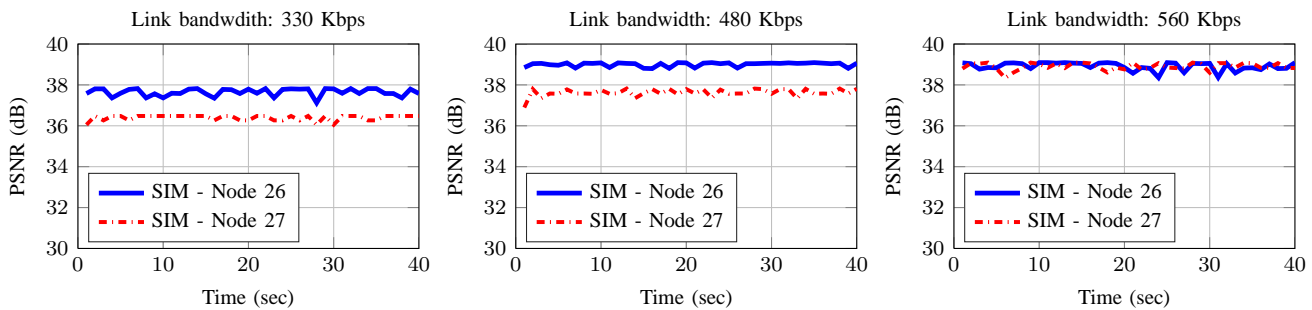

Fig. \ref{fig:quality_vs_time} depicts the average PSNR of the delivered video at nodes 26 and 27 as a function of time. The playback delay is set to 1000msec. We can see that our network coding enabled NDN protocol achieves low jitter in the video quality. The quality of the decoded video remains close to the value that is expected based on the optimal rate allocation. The results indicate that the majority of generations are decoded correctly with only a few of them decoded at lower quality or not decoded at all. This is true even when the bandwidth resources get scarce (leftmost figure in Fig. \ref{fig:quality_vs_time}). Finally, it is worth noting that the employed forwarding strategy ensures that the probability of receiving a redundant packet is zero and non-innovative packets can be received only due to the randomness of the network coding operations.




\section{Conclusions}
\label{sec:discussion}

In this paper, we have presented a novel network coding enabled NDN architecture for the delivery of scalable video. In order to design the optimal content-aware forwarding strategy, we have formulated a rate allocation optimization problem which decides on the optimal rates of Interest messages sent by clients and intermediate nodes. This rate allocation ensures that the achieved flow of Data objects maximizes the average quality of the video delivered to the client population. We have also proposed the necessary modification to the standard NDN architecture in order to support the handling of Interest messages and Data objects when intermediate nodes perform network coding. In order to resolve the practical issues that are caused by the use of network coding, we have proposed the use of Bloom filters that compactly store additional information about the Interest messages and Data objects. We have evaluated our protocol in terms of the achieved video quality for the delivery of scalable video. The results indicate that the proposed architecture achieves a close to optimal performance.

Our future work will include the extension of the proposed architecture to the cases where link losses affect the delivery of Interest messages and Data packets. In this case, appropriate mechanisms need to be designed in order to properly handle the re-transmission of Interest and Data packets. We will also investigate the design of distributed/heuristic algorithms for the forwarding of Interest messages that will be able to guarantee a close to optimal performance of our system without the need for central control.




\bibliographystyle{IEEEtran}

\begin{thebibliography}{10}
\providecommand{\url}[1]{#1}
\csname url@samestyle\endcsname
\providecommand{\newblock}{\relax}
\providecommand{\bibinfo}[2]{#2}
\providecommand{\BIBentrySTDinterwordspacing}{\spaceskip=0pt\relax}
\providecommand{\BIBentryALTinterwordstretchfactor}{4}
\providecommand{\BIBentryALTinterwordspacing}{\spaceskip=\fontdimen2\font plus
\BIBentryALTinterwordstretchfactor\fontdimen3\font minus
  \fontdimen4\font\relax}
\providecommand{\BIBforeignlanguage}[2]{{%
\expandafter\ifx\csname l@#1\endcsname\relax
\typeout{** WARNING: IEEEtran.bst: No hyphenation pattern has been}%
\typeout{** loaded for the language `#1'. Using the pattern for}%
\typeout{** the default language instead.}%
\else
\language=\csname l@#1\endcsname
\fi
#2}}
\providecommand{\BIBdecl}{\relax}
\BIBdecl

\bibitem{Cisco:15}
``{Cisco Visual Networking Index-Forecast and Methodology, 2014Ð2019},''
  White paper, May 2015.

\bibitem{SVC2005}
{ITU-T and ISO/IEC JTC 1}, ``{Advanced Video Coding for Generic Audiovisual
  Sevices, Amendment 3: Scalable Video Coding},'' \emph{Draft {ITU-T}
  Recommnedation {H.264 - ISO/IEC 14496-10 (AVC)}}, Apr. 2005.

\bibitem{SodagarMM2011}
I.~Sodagar, ``{The MPEG-DASH Standard for Multimedia Streaming Over the
  Internet},'' \emph{IEEE Multimedia}, vol.~18, no.~4, pp. 62--67, Apr. 2011.

\bibitem{StockhammerMMSys2011}
T.~Stockhammer, ``{Dynamic Adaptive Streaming over HTTP: Standards and Design
  Principles},'' in \emph{Proc. of the 2nd Annual ACM Conf. on Multimedia
  Systems}, San Jose, CA, USA, Feb. 2011, pp. 133--144.

\bibitem{XylomenosComSurvey2014}
G.~Xylomenos, C.~N. Ververidis, V.~A. Siris, N.~Fotiou, C.~Tsilopoulos,
  X.~Vasilakos, K.~V. Katsaros, and G.~C. Polyzos, ``{A Survey of
  Information-Centric Networking Research},'' \emph{IEEE Communication Surveys
  and Tutorials}, vol.~16, no.~2, pp. 1024--1048, Apr. - June 2014.

\bibitem{JacobsonCoNEXT2009}
V.~Jacobson, D.~K. Smetters, J.~D. Thornton, M.~F. Plass, N.~H. Briggs, and
  R.~L. Braynard, ``{Networking Named Content},'' in \emph{Proc. of ACM
  CoNEXT}, Rome, Italy, Dec. 2009, pp. 1--12.

\bibitem{TsilopoulosPV2013}
C.~Tsilopoulos, G.~Xylomenos, and G.~C. Polyzos, ``{Are Information-Centric
  Networks Video-Ready?}'' in \emph{Proc. of IEEE Packet Video Workshop,
  PV'13}, San Jose, CA, USA, Dec. 2013, pp. 1--8.

\bibitem{ByunICC2013}
D.~Byun, B.-J. Lee, and M.-W. Jang, ``{Adaptive Flow Control via Interest
  Aggregation in CCN},'' in \emph{Proc. of Int. Conf. on Communications,
  ICC'13}, Budapest, Hungary, June 2013.

\bibitem{TsilopoulosACMICN2011}
C.~Tsilopoulos and G.~Xylomenos, ``{Supporting Diverse Traffic Types in
  Information Centric Networks},'' in \emph{Proc. of ACM SIGCOMM ICN Workshop},
  Toronto, ON, Canada, Aug. 2011, pp. 13--18.

\bibitem{LiuICC2013}
Y.~Liu, J.~Geurts, J.~Point, S.~Lederer, B.~Rainer, C.~Mueller, C.~Timmerer,
  and H.~Hellwagner, ``{Dynamic Adaptive Streaming over CCN: A Caching and
  Overhead Analysis},'' in \emph{Proc. of IEEE Int. Conf. on Communications,
  ICC'2013}, Budapest, Hungary, June 2013.

\bibitem{MontpetitNoM2012}
M.-J. Montpetit, C.~Westphal, and D.~Trossen, ``{Network Coding Meets
  Information-centric Networking: an Architectural Case for Information
  Dispersion Through Native Network Coding},'' in \emph{Proc. of ACM Workshop
  on Emerging Name-Oriented Mobile Networking Design- Architecture, Algorithms
  and Applications,}, Hilton Head Island, SC, USA, June 2012.

\bibitem{AhlswedeTIT2000}
R.~Ahlswede, N.~Cai, S.-Y.~R. Li, and R.~W. Yeung, ``{Network Information
  Flow},'' \emph{IEEE Trans. Information Theory}, vol.~46, no.~4, pp.
  1204--1216, July 2000.

\bibitem{ThomosTMM2011}
N.~Thomos, J.~Chakareski, and P.~Frossard, ``{Prioritized Distributed Video
  Delivery With Randomized Network Coding},'' \emph{IEEE Trans. Multimedia},
  vol.~13, no.~4, pp. 776--787, Aug. 2011.

\bibitem{HoAllerton2003}
T.~Ho, M.~M\'edard, J.~Shi, M.~Effros, and D.~R. Karger, ``{On Randomized
  Network Coding},'' in \emph{Proc. of the 41st Allerton Conf. on
  Communication, Control and Computing}, Monticello, IL, USA, Oct. 2003.

\bibitem{ChouAllerton2003}
P.~A. Chou, Y.~Wu, and K.~Jain, ``{Practical Network Coding},'' in \emph{Proc.
  of the 41st Allerton Conf. on Communication, Control and Computing},
  Monticello, IL, USA, Oct. 2003.

\bibitem{WuACMICN2013}
Q.~Wu, Z.~Li, and G.~Xie, ``{CodingCache: Multipath-aware CCN Cache with
  Network Coding},'' in \emph{Proc. of ACM SIGCOMM ICN Workshop}, Hong Kong,
  China, Aug. 2013, pp. 41--42.

\bibitem{LloreaICC2013}
J.~Llorea, A.~Tulino, K.~Guan, and D.~Kilper, ``{Network-Coded Caching-Aided
  Multicast for Efficient Content Delivery},'' in \emph{Proc. of IEEE Int.
  Conf. on Communications, ICC'2013}, Budapest, Hungary, June 2013, pp.
  3557--3562.

\bibitem{AnastasiadesICC2015}
C.~Anastasiades, N.~Thomos, A.~Striffeler, and T.~Braun, ``{RC-NDN: Raptor
  Codes Enabled Named Data Networking},'' in \emph{Proc. of IEEE Int. Conf. on
  Communications, ICC'2015}, London, UK, 2015.

\bibitem{KurdogluICME2011}
E.~Kurdoglu, N.~Thomos, and P.~Frossard, ``{Scalable Video Dissemination with
  Prioritized Network Coding},'' in \emph{Proc. of Streaming and Media
  Communication Workshop, StreamComm'11 (in conjunction with ICME'11)},
  Barcelona, Spain, July 2011.

\bibitem{LiINFOCOM05}
Z.~Li, B.~Li, D.~Jiang, and L.~C. Lau, ``{On Achieving Optimal Throughput With
  Network Coding},'' in \emph{Proc of 24th INFOCOM}, Miami, FL, USA, Mar. 2005,
  pp. 2184--2194.

\bibitem{Sherali96}
H.~D. Sherali and G.~Choi, ``{Recovery of Primal Solutions When Using
  Subgradient Optimization Methods to Solve Lagrangian Duals of Linear
  Programs},'' \emph{{Elsevier Operations Research Letters}}, vol.~19, no.~3,
  pp. 105--113, Sep. 1996.

\bibitem{Bloom70}
B.~H. Bloom, ``{Space/Time Trade-offs in Hash Coding with Allowable Errors},''
  \emph{{Communications of the ACM}}, vol.~13, no.~7, pp. 422--426, Jul. 1970.

\bibitem{Planetlab2015}
``{P}lanet{L}ab,'' https://www.planet-lab.org/.

\bibitem{ClejuTMM2011}
N.~Cleju, N.~Thomos, and P.~Frossard, ``{Selection of Network Coding Nodes for
  Minimal Playback Delay in Streaming Overlays},'' \emph{IEEE Trans.
  Multimedia}, vol.~13, no.~5, pp. 1103--1115, Oct. 2011.

\bibitem{ns3}
``The network simulator - ns3,'' http://www.nsnam.org/.

\end{thebibliography}

\end{document}